\documentclass[twocolumn]{aastex63}
\usepackage{booktabs}
\usepackage{natbib}
\usepackage{multirow}

\shorttitle{Flash ionization signatures in type Ibn supernova SN~2019uo}
\shortauthors{Gangopadhyay et al.}

\begin{document}

\title{Flash ionization signatures in the type Ibn supernova SN~2019uo}

\correspondingauthor{Anjasha Gangopadhyay}
\email{anjasha@aries.res.in, anjashagangopadhyay@gmail.com}

\author[0000-0002-3884-5637]{Anjasha Gangopadhyay}
\affiliation{Aryabhatta Research Institute of observational sciencES, Manora Peak, Nainital 263 002 India}
\affiliation{School of Studies in Physics and Astrophysics, Pandit Ravishankar Shukla University, Chattisgarh 492 010, India}

\author[0000-0003-1637-267X]{Kuntal Misra}
\affiliation{Aryabhatta Research Institute of observational sciencES, Manora Peak, Nainital 263 002 India}
\affiliation{Department of Physics, University of California, 1 Shields Ave, Davis, CA 95616-5270, USA}

\author{Daichi Hiramatsu}
\affiliation{Las Cumbres Observatory, 6740 Cortona Drive Suite 102, Goleta, CA, 93117-5575 USA}
\affiliation{Department of Physics, University of California, Santa Barbara, CA 93106-9530, USA}

\author{Shan-Qin Wang}
\affiliation{Guangxi Key Laboratory for Relativistic Astrophysics, School of Physical Science and Technology, Guangxi University, Nanning 530004, People's Republic of China}

\author{Griffin Hosseinzadeh}
\affiliation{Center for Astrophysics \textbar{} Harvard $\&$ Smithsonian, 60 Garden Street, Cambridge, MA 02138-1516, USA}

\author{Xiaofeng Wang}
\affiliation{Physics Department and Tsinghua Center for Astrophysics, Tsinghua University, China}

\author{Stefano Valenti}
\affiliation{Department of Physics, University of California, 1 Shields Ave, Davis, CA 95616-5270, USA}

\author{Jujia Zhang}
\affiliation{Yunnan Astronomical Observatory of China, Chinese Academy of Sciences, Kunming, 650011, China}

\author{D. Andrew Howell}
\affiliation{Las Cumbres Observatory, 6740 Cortona Drive Suite 102, Goleta, CA, 93117-5575 USA}
\affiliation{Department of Physics, University of California, Santa Barbara, CA 93106-9530, USA}

\author[0000-0001-7090-4898]{Iair Arcavi}
\affiliation{The School of Physics and Astronomy, Tel Aviv University, Tel Aviv 69978, Israel}
\affiliation{CIFAR Azrieli Global Scholars program, CIFAR, Toronto, Canada}

\author[0000-0003-3533-7183]{G.C. Anupama}
\affiliation{Indian Institute of Astrophysics, Koramangala 2nd Block, Bangalore 560034, India}

\author{Jamison Burke}
\affiliation{Las Cumbres Observatory, 6740 Cortona Drive Suite 102, Goleta, CA, 93117-5575 USA}
\affiliation{Department of Physics, University of California, Santa Barbara, CA 93106-9530, USA}

\author[0000-0001-6191-7160]{Raya Dastidar}
\affiliation{Aryabhatta Research Institute of observational sciencES, Manora Peak, Nainital 263 002 India}
\affiliation{Department of Physics $\&$ Astrophysics, University of Delhi, Delhi-110 007}

\author{Koichi Itagaki}
\affiliation{Itagaki Astronomical Observatory, Japan}

\author[0000-0001-7225-2475]{Brajesh Kumar}
\affiliation{Indian Institute of Astrophysics, Koramangala 2nd Block, Bangalore 560034, India}

\author{Brijesh Kumar}
\affiliation{Aryabhatta Research Institute of observational sciencES, Manora Peak, Nainital 263 002 India}

\author{Long Li}
\affiliation{Guangxi Key Laboratory for Relativistic Astrophysics, School of Physical Science and Technology, Guangxi University, Nanning 530004, People's Republic of China}

\author{Curtis McCully}
\affiliation{Las Cumbres Observatory, 6740 Cortona Drive Suite 102, Goleta, CA, 93117-5575 USA}
\affiliation{Department of Physics, University of California, Santa Barbara, CA 93106-9530, USA}

\author{Jun Mo}
\affiliation{Physics Department and Tsinghua Center for Astrophysics, Tsinghua University, China}

\author{Shashi Bhushan Pandey}
\affiliation{Aryabhatta Research Institute of observational sciencES, Manora Peak, Nainital 263 002 India}

\author{Craig Pellegrino}
\affiliation{Las Cumbres Observatory, 6740 Cortona Drive Suite 102, Goleta, CA, 93117-5575 USA}
\affiliation{Department of Physics, University of California, Santa Barbara, CA 93106-9530, USA}

\author{Hanna Sai}
\affiliation{Physics Department and Tsinghua Center for Astrophysics, Tsinghua University, China}

\author{D.K. Sahu}
\affiliation{Indian Institute of Astrophysics, Koramangala 2nd Block, Bangalore 560034, India}

\author{Pankaj Sanwal}
\affiliation{Aryabhatta Research Institute of observational sciencES, Manora Peak, Nainital 263 002 India}
\affiliation{School of Studies in Physics and Astrophysics, Pandit Ravishankar Shukla University, Chattisgarh 492 010, India}

\author[0000-0003-2091-622X]{Avinash Singh}
\affiliation{Indian Institute of Astrophysics, Koramangala 2nd Block, Bangalore 560034, India}
\affiliation{Joint Astronomy Programme, Department of Physics, Indian Institute of Science, Bengaluru 560012, India}

\author{Mridweeka Singh}
\affiliation{Korea Astronomy and Space Science Institute, 776 Daedeokdae-ro, Yuseong-gu, Daejeon 34055, Republic of Korea}
\affiliation{Aryabhatta Research Institute of observational sciencES, Manora Peak, Nainital 263 002 India}

\author{Jicheng Zhang} 
\affiliation{Physics Department and Tsinghua Center for Astrophysics, Tsinghua University, China}

\author{Tianmeng Zhang}
\affiliation{National Astronomical Observatory of China, Chinese Academy of Sciences, Beijing, 100012, China}

\author{Xinhan Zhang}
\affiliation{Physics Department and Tsinghua Center for Astrophysics, Tsinghua University, China}



\begin{abstract}

We present photometric and spectroscopic observations of the type Ibn supernova (SN) 2019uo, the second ever SN~Ibn with flash ionization (\ion{He}{2}, \ion{C}{3}, \ion{N}{3}) features in its early spectra. SN~2019uo displays a rapid post-peak luminosity decline of 0.1 mag d$^{-1}$ similar to most of the SNe~Ibn, but is fainter ($M^V_{max} = -18.30 \pm 0.24$\ mag) than a typical SN~Ibn and shows {\bf a} color evolution that places it between SNe~Ib and the most extreme SNe~Ibn. SN~2019uo shows P-cygni \ion{He}{1} features in the early spectra which gradually evolves and becomes emission dominated post peak. It also shows faster evolution in line velocities as compared to most other members of the type Ibn subclass. The bolometric light curve is fairly described by a $^{56}$Ni + circumstellar interaction model.

\end{abstract}


\section{Introduction}
\label{1}
Supernovae (SNe) undergoing interaction with a circumstellar medium (CSM) provide a unique window in the evolutionary phases of stars. Interaction, in general, produces narrow emission lines --- broader than \ion{H}{2} regions but narrower than lines arising from the outer ejecta of the supernova \citep{2017ApJ...836..158H}. However, in some cases interaction happens below the photosphere without any observable narrow emission lines \citep[e.g.][]{2017ApJ...838...28M,2018MNRAS.477...74A}. SNe~IIn \citep{1990MNRAS.244..269S} and SNe~Ia-CSM display narrow H lines indicative of interaction with a H-rich CSM. Approximately 1\% of core-collapse SNe (CCSNe) show little H and narrow He features ($\sim$2000~km~s$^{-1}$). With the discovery of SN~2006jc, \cite{2007Natur.447..829P} introduced this class as SNe~Ibn, whose spectral features show interaction signatures between SN ejecta and a He-rich CSM. This is defined in analogy with SNe~IIn, which show narrow H features  \citep{1990MNRAS.244..269S}.
SNe that are embedded in dense CSM may also show short-lived narrow high ionization emission lines ($\leq$10~days) owing to the recombination of the CSM following the shock breakout flash. These features are known as ``flash features" 
\citep[e.g.][]{2014Natur.509..471G}.
\cite{2017ApJ...836..158H} analysed a sample of SN~Ibn light curves and showed that unlike SNe~IIn, SNe~Ibn are rather uniform in their light curve shape with rapid decay rates of 0.05--0.15~mag~d$^{-1}$. SNe~Ibn may have double-peaked light curves like SNe~IIn, but they show a faster rise than SNe~IIn \citep{2017ApJ...836..158H}. On the other hand, \cite{2016MNRAS.456..853P} showed that the class is heterogeneous with many outliers: OGLE-2012-SN-006 \citep{2015MNRAS.449.1941P} has a very slow decline; LSQ13ccw \citep{2015MNRAS.449.1954P} is faint and fast-declining; SNe~2005la and 2011hw \citep{2015MNRAS.449.1921P} are transitional type~IIn/Ibn events; SN~2010al \citep{2015MNRAS.449.1921P} is the earliest detected SN~Ibn with a slow rise and decline. \cite{2019arXiv191006016K} recently identified a rapid evolving SN 2018bcc. SNe~Ibn have bluer continuum than other CCSNe. Some SNe~Ibn show P~Cygni \ion{He}{1} emission, while others transition from narrow to intermediate-width \ion{He}{1} emissions \citep{2017ApJ...836..158H}. 

So far, only indirect progenitor constraints for SNe~Ibn are available. 
\cite{2007Natur.447..829P} suggest Wolf-Rayet (WR) H-free atmospheres generate the He-rich CSM. The best studied case for unstable mass loss from a WR progenitor is SN~2006jc, for which an optical transient was detected at the SN location two years prior to explosion \citep{2007ApJ...657L.105F,2007Natur.447..829P,2008ApJ...680..568S}. Alternatively, CSM can be produced by stripping material from envelopes of massive binaries \citep{2007ApJ...657L.105F}. However, a low-mass progenitor has been suggested for PS1-12sk, which occurred in a non-star-forming host \citep{2013ApJ...769...39S,2019ApJ...871L...9H} --- unlikely for a CCSN \citep[$\leq$0.2\%;][]{2012A&A...544A..81H}. A very recent study by \cite{2019arXiv190907999S} for SNe~2006jc and 2015G implies an interacting binary progenitor scenario, based on late time UV/optical {\it HST} images.

In this paper we study the evolution of one such type Ibn SN~2019uo which was discovered on 2019 January 17.8 UT (JD 2458501.3) by Koichi Itagaki at $\mathrm{R.A.} = 12^\mathrm{h}02^\mathrm{m}36.5^\mathrm{s}$, $\mathrm{Decl.} = +41\degr03'42''$ (J2000.0). The SN location is 0\farcs4 east and 27\farcs2 north of the center of the galaxy UGC~7020 at a redshift of 0.020454 \citep{2019ATel12410....1Z}. SN~2019uo was classified on 2019 January 19.9 UT as a SN~II \citep{2019ATel12410....1Z} with the spectrum obtained with the Yunnan Faint Object Spectrograph and Camera (YFOSC) mounted on the 2.4~m LiJiang Telescope (LJT) at Yunnan Observatory (YNAO). \cite{2019ATel12410....1Z} reported that the spectrum depicted a blue continuum and highly ionized ``flash features" such as \ion{N}{5}, \ion{He}{2} and \ion{O}{5}. However, this classification of type II SN was modified later by \cite{2019TNSCR.188....1F} and SN 2019uo was classified as a type Ibn. Prominent narrow emission lines of \ion{He}{1} in the initial spectra of SN~2019uo indicating a P-cygni velocity of 650 km~s$^{-1}$ justified the type Ibn classification. SN~2019uo is the second SN~Ibn to show these features after SN~2010al. Adopting $H_0 = 73$~km~s$^{-1}$~Mpc$^{-1}$, we obtain a luminosity distance of 88.8~Mpc for SN~2019uo. The Milky Way extinction along the line of sight of SN~2019uo is $A_V = 0.035$~mag \citep{2011ApJ...737..103S}. For estimating the extinction along the line of sight within host galaxy, we estimate equivalent widths of the \ion{Na}{1}D line in the first three spectra of SN~2019uo. Using the formulation by \cite{1997A&A...318..269M} and \cite{2012MNRAS.426.1465P}, we estimate $A_V = 0.2517$~mag. This estimate also brings the $B-V$ colors of SN~2019uo into close agreement with SNe~2006jc and 2010al. Thus, we adopt a total $A_V = 0.287$~mag. The temporal and spectral evolution of SN 2019uo and the detailed modeling of the bolometric light curve is discussed in the sections to follow. 
\begin{table*}
\centering
\caption{Photometry of SN~2019uo} 
\resizebox{\hsize}{!}{\begin{tabular}{c c c c c c c c c c c}
\toprule
Date  & JD & Phase$^{\dagger}$       & 		$U$             & $B$             & $g$              & $V$                 & $r$              &  $i$         & Telescope\\    
(yyyy-mm-dd) & (2458000+)   & (day)   & 		(mag)           & (mag)           & (mag)            & (mag)               & (mag)            & (mag)        &            \\
\midrule
2019-01-18 & 501.8 & -6.8  & ---  &  17.865$\pm$0.107 & 17.637$\pm$0.097 & 17.927$\pm$0.146 & 17.936$\pm$0.139 & --- & LCO \\
2019-01-20 & 503.9 & -4.7  & 16.452$\pm$0.033  & 16.937$\pm$0.027 & 16.785$\pm$0.013 & 17.052$\pm$0.026 & 17.039$\pm$0.019 & 17.353$\pm$0.033 & LCO \\
2019-01-20 & 504.3 & -4.3  & ---  &  16.817$\pm$0.052  & 16.908$\pm$0.162  & 17.262$\pm$0.050  &    --- & 17.423$\pm$0.052   & TNT \\
2019-01-21 & 505.3 & -3.3  & ---  &  16.623$\pm$0.013  & 16.768$\pm$0.018  & 17.014$\pm$0.030  &  16.901$\pm$0.029 & 17.328$\pm$0.038 & TNT \\  
2019-01-21 & 505.2 & -3.4  & ---  &  ---               & ---               & 16.882$\pm$0.08   &  ---             & ---             & 0.7m \\
2019-01-22 & 506.2 & -2.4  & ---  & ---                & ---	           & 16.777$\pm$0.063  &  ---             & ---           & 0.7m \\ 
2019-01-23 & 506.8 & -1.8  & 16.122$\pm$0.074  & ---   & 16.613$\pm$0.039  & 16.883$\pm$0.051  &  ---             & ---              & LCO \\
2019-01-23 & 507.3 & -1.3  & ---  &  16.654$\pm$0.036  & 16.699$\pm$0.022  & 16.991$\pm$0.034  &  16.821$\pm$0.034  & 17.187$\pm$0.042   & LCO \\
2019-01-24 & 507.8 & -0.8  & 16.252$\pm$0.075 & 16.731$\pm$0.031   & 16.532$\pm$0.015  & 16.669$\pm$0.027  &  16.676$\pm$0.022  & 17.215$\pm$0.022 & LCO \\
2019-01-25 & 508.1 & -0.6  & ---  & ---       & ---    & 16.673$\pm$0.024 & ---   &  17.037$\pm$0.015 & LJT   \\
2019-01-25 & 508.3 & -0.3  & 16.015$\pm$0.020 & 16.825$\pm$0.031 & 16.655$\pm$0.019 & 16.616$\pm$0.101 & 16.782$\pm$0.022 & 17.027$\pm$0.0272 & LJT,TNT,0.7m \\
2019-01-25 & 508.8 & 0.1   & 16.199$\pm$0.051  & 16.740$\pm$0.026   & 16.565$\pm$0.017 & 16.685$\pm$0.028 & 16.654$\pm$0.023 & 17.003$\pm$0.055  & LCO \\
2019-01-28   & 511.4 & 2.7   & ---        & --- & --- & 16.732$\pm$0.143 & --- & --- & 0.7m \\
2019-01-28   & 511.8 & 3.1   & 16.331$\pm$0.028  & 17.156$\pm$0.017  & ---               & 16.749$\pm$0.0153 &   16.646$\pm$0.010  & ---                & DFOT   \\
2019-01-30   & 513.1 & 4.5   & --- & --- & --- & 16.786$\pm$0.085  &    ---   & ---                & ST \\
2019-02-01   & 515.7 & 7.1   & 17.334$\pm$0.049 & 17.447$\pm$0.026  & 17.281$\pm$0.009  & 17.307$\pm$0.019  &    17.237$\pm$0.011  & 17.430$\pm$0.021   & LCO \\
2019-02-02   & 516.3 & 7.6   &   --- &	17.547$\pm$0.016  & 17.257$\pm$0.011  & 17.655$\pm$0.022  & 17.532$\pm$0.016  & 17.696$\pm$0.013   & TNT \\
2019-02-04   & 518.4 & 9.8   &   --- &  18.275$\pm$0.016  & ---      &  ---   & 17.847$\pm$0.017  &  17.948$\pm$0.014 & ST \\
2019-02-04   & 518.8 & 10.2  &   18.262$\pm$0.082  &	18.276$\pm$0.029  & 17.921$\pm$0.016  & 17.882$\pm$0.025  &    17.835$\pm$0.021  & 17.938$\pm$0.030   & LCO \\
2019-02-05   & 519.2 & 10.5  &  ---   & 18.754$\pm$0.031& ---               & 17.885$\pm$0.024  &    17.856$\pm$0.017  & ---                & ST \\
2019-02-06   & 520.3 & 11.7  &  ---   &	18.862$\pm$0.024  & ---               & 18.141$\pm$0.025  &    18.479$\pm$0.027  & 18.592$\pm$0.033   & ST \\
2019-02-08   & 523.4 & 14.8  & 19.787$\pm$0.041 & 19.058$\pm$0.027 & ---               & 18.316$\pm$0.031  &    18.900$\pm$0.001  & 18.808$\pm$0.052   & HCT \\
2019-02-09   & 524.4 & 15.7  & ---  & 19.224$\pm$0.069  & --- & 18.598$\pm$0.029  &   18.911$\pm$0.037 & 19.091$\pm$0.033  & ST \\
2019-02-12   & 527.0 & 18.4  & ---  & 19.487$\pm$0.053  & 19.169$\pm$0.037  & 18.946$\pm$0.043  &    19.012$\pm$0.048 & 19.136$\pm$0.075   & LCO \\ 
2019-02-20   & 535.2 & 26.6   &	     ---	      &	---	       	  & 20.037$\pm$0.218  & ---               &    ---               & 20.287$\pm$0.115   & TNT \\
2019-02-21   & 536.2 & 27.6   &	     ---	      &	---	       	  & 20.228$\pm$0.214  & ---               &    ---               &   ---              & TNT \\
2019-02-23   & 538.1 &  29.5   &	     ---	      &	---	       	  & 20.265$\pm$0.312  & ---               &    ---               &   ---              & TNT \\
2019-03-02   & 544.3 & 36.2   &	     --- & 21.231$\pm$0.274  & 20.726$\pm$0.138  & 20.526$\pm$0.178  &    21.032$\pm$0.305  & 21.096$\pm$0.215   & LCO \\
\bottomrule
\multicolumn{4}{@{}l}{$^\dagger$ with respect to $\mathrm{JD_{max}} = 2458508.65$.}
\end{tabular}}
\label{tab:observation_log_2019uo}     
\end{table*}

\section{Data Acquisition and Reduction}
\label{2}

We observed SN~2019uo with Las Cumbres Observatory (LCO) in the \textit{UBVgri} filters from $\sim$2 to 106 days after discovery. Augmenting the LCO data, photometric observations in {\it UBVRI/ugri} were also taken with 0.7m BITRAN-CCD Imaging System located in Japan; 0.8m Tsinghua-NAOC Telescope (TNT), Xinglong Observatory, China; 1.04m Sampurnanand Telescope (ST); 1.30m Devasthal Fast Optical Telescope (DFOT), ARIES, India; 2.00m Himalayan Chandra Telescope (HCT), IAO, Hanle, India and Lijiang 2.4m Telescope (LJT), Yunnan Observatories (YNAO), China. We performed image subtraction using High Order Transform of PSF ANd Template Subtraction (HOTPANTS)\footnote{\url{https://github.com/acbecker/hotpants}}\citep{2015ascl.soft04004B}. The instrumental magnitudes were estimated using IRAF\footnote{Image Reduction and Analysis Facility}\citep{1986SPIE..627..733T,1993ASPC...52..173T} and DAOPHOT\footnote{Dominion Astrophysical Observatory Photometry}\citep{1987PASP...99..191S}. The LCO photometry was done using \texttt{lcogtsnpipe}\footnote{\url{https://github.com/svalenti/lcogtsnpipe}} (see \citealp{2011MNRAS.416.3138V,2016MNRAS.459.3939V}) on the difference images. The instrumental SN magnitudes were calibrated using the standard magnitudes of a number of local  stars in the SN field obtained from the Sloan Digital Sky Survey (SDSS) catalog for the {\it gri} bands and the Landolt standard fields taken on the same night by the same instrument as the science images for {\it UBV}. Wherever required, the {\it RI} magnitudes were converted to {\it ri} using the equations of \cite{2006A&A...460..339J}. The photometry of SN~2019uo is presented in Table \ref{tab:observation_log_2019uo}.
\begin{table*}
\caption{Log of spectroscopic observations of SN~2019uo. \label{tab:2019uo_spec_obs}}
\begin{center}
\smallskip
\footnotesize\addtolength{\tabcolsep}{-2pt}
\resizebox{\textwidth}{!}{
\begin{tabular}{c c c c c c}
\toprule
\colhead{Date} & \colhead{$\mathrm{JD} - 2458000$} & \colhead{Phase$^{\dagger}$}  &    \colhead{Telescope}  &     \colhead{Instrument}  &		\colhead{Range (\AA)} \\
\midrule
2019-01-19 & 503.4 & -5.2      &  2.4~m LJT    &    YFOSC                  &   3500-8800           \\
2019-01-20 & 503.9 & -4.7      &  2.0~m FTN    &    FLOYDS 	          &   3200-9000   \\
2019-01-21 & 504.9 & -3.7      &   2.0~m FTN    &    FLOYDS                 &   3200-9000             \\
2019-01-23 & 506.9 & -1.7      &   2.0~m FTN    &    FLOYDS                 &   3200-9000   \\
2019-01-24 & 508.3 & -0.3     &   2.4~m LJT    &    YFOSC                  &   3500-8800 \\
2019-01-28 & 512.4 & 3.8      &   2.0~m FTN    &    FLOYDS                 &   3200-9000  \\
2019-02-05 & 519.9 & 11.3     &    2.0~m FTN    &    FLOYDS                 &   3200-9000   \\
2019-02-08 & 523.8 & 15.2     &   2.2~m China  &    BFOSC       &   4000-10000\\
2019-02-14 & 529.2 & 20.6     &   2.4~m LJT    &    YFOSC                  &   3500-8800 \\
\bottomrule
\multicolumn{4}{@{}l}{$^\dagger$ with respect to $\mathrm{JD_{max}} = 2458508.65$.}
\end{tabular}}
\label{tab:spectra_log_2019uo}  
\end{center}
\end{table*}          

\begin{figure}
	\begin{center}
		\includegraphics[width=0.5\textwidth]{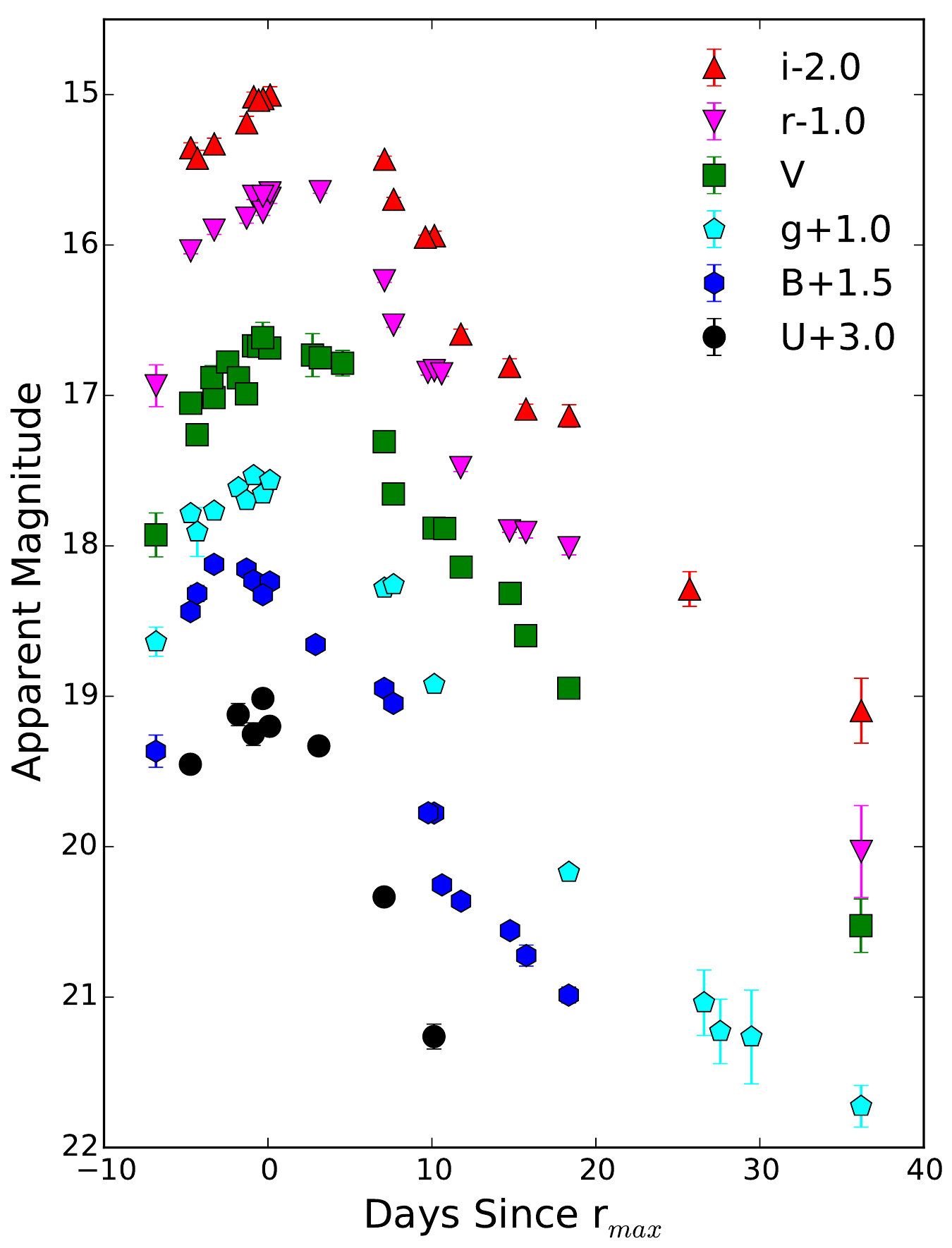}
	\end{center}
	\caption{{\it UBVgri} light curve evolution of SN~2019uo.}
	\label{fig:lc_2019uo}
\end{figure}

The spectroscopic observations were taken at 9 epochs spanning up to $\sim$88 days after discovery. The 1D wavelength- and flux-calibrated spectra were extracted using the floydsspec pipeline \citep{2014MNRAS.438L.101V} for the LCO data. Spectroscopic data reduction of the 2.2~m and 2.4~m telescopes was done using the APALL task in IRAF followed by wavelength and flux calibration. The slit loss corrections were done by scaling the spectra to the photometry. Finally, the spectra were corrected for the heliocentric redshift of the host galaxy. The log of spectroscopic observations is given in Table \ref{tab:spectra_log_2019uo}. 

\section{Photometric Evolution of SN~2019uo}
\label{3}
The complete multi-band light curve of SN~2019uo is shown in Figure~\ref{fig:lc_2019uo}. With our available observations, we were able to trace the epoch of maximum in all the bands. The date of maximum and its brightness were determined by fitting a cubic spline to the {\it UBVgri} light curves. The maximum in $r$-band occurred on JD $2458508.6 \pm 0.5$ at an apparent magnitude of $16.66 \pm 0.03$~mag. The errors reported are obtained from interpolated measurements around the peak. We use days since $r$-maximum ($r_\mathrm{max}$) as a reference epoch throughout the paper. Assuming that the discovery date is close to explosion, we estimated a rise time of $8.7 \pm 1.3$~days. This is similar to iPTF14aki and iPTF15akq (c.f.\ Table 4; \citealp{2017ApJ...836..158H}). 

The $r$-band light curve, between 0--36~days, decays with a rate of $0.126 \pm 0.005$~mag~d$^{-1}$. The $g$, $B$, $V$ and $i$ bands follow approximately the same decline rate. The sample of SNe~Ibn in \cite{2017ApJ...836..158H} are fast-evolving with a typical decline rate of 0.1~mag~d$^{-1}$ during the first month post-maximum. SN~2019uo follows the same decline rate. 

\begin{figure}
	\begin{center}
		\includegraphics[width=0.5\textwidth]{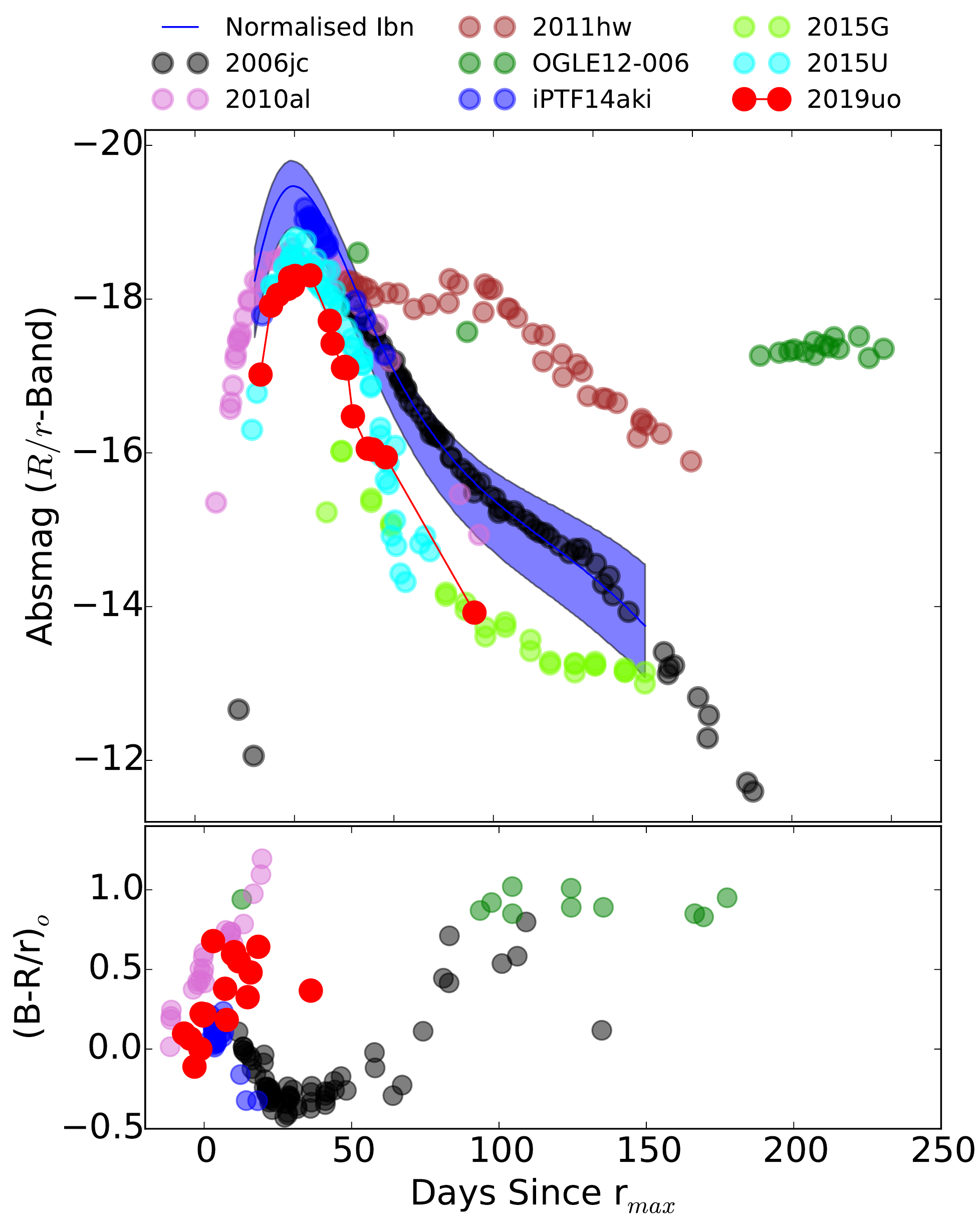}
	\end{center}
	\caption{$R/r$-band absolute magnitude light curve and $B-R/r$ colour curve of SN~2019uo. The comparison sample includes SNe~2006jc \citep{2007Natur.447..829P,2007ApJ...657L.105F}, 2010al \citep{2015MNRAS.449.1954P}, OGLE-SN-006 \citep{2015MNRAS.449.1941P}, 2011hw \citep{2015MNRAS.449.1954P}, iPTF14aki \citep{2017ApJ...836..158H}, 2015U \citep{2016MNRAS.461.3057S,2017ApJ...836..158H} and 2015G \citep{2017ApJ...836..158H}.}
	\label{fig:absmag}
\end{figure}

Figure~\ref{fig:absmag} shows the absolute magnitude light curve of SN~2019uo along with other SNe~Ibn after correcting for distance and extinction. The peak $r$-band absolute magnitude of SN~2019uo is $-18.30 \pm 0.24$~mag, which is at the fainter end of SN~Ibn sample. The blue band in Figure~\ref{fig:absmag} shows the average light curve (comprising of 95\% of the SN~Ibn data) of SNe~Ibn taken from \cite{2017ApJ...836..158H}. The average light curve was generated by using a Gaussian process to fit a smooth curve to the combined light curves on the sample of \cite{2017ApJ...836..158H}. The fit was performed in log-log space to ensure consistency and smoothness between the early and late time light curves. The average light curve, thus, generated also uses the Gaussian process to fit positive and negative residuals. It is to note that SN~2019uo is $\sim$1.2~mag fainter than the normalized SNe~Ibn light curve.

We compare the $B-R/r$ color evolution of SN~2019uo with a number of type Ibn SNe, which usually show heterogeneity in their color evolution. The $B-r$ color of SN~2019uo increases up to 0.64~mag $\sim$20 days post $r_{max}$, subsequently becoming blue at $\sim$36~days. Similarly, for SN~2010al and iPTF14aki the $B-r$ color increases up to $\sim$1~mag, $\sim$30 days post $R_{max}$. Thus, SN~2019uo shows a color evolution similar to SN~2010al and iPTF14aki. At similar epochs, the color evolution of SN~2006jc was extremely blue ($-0.5$~mag). SN~2006jc, then shows an overall flatter color evolution. The early blue colour are typical of type Ibn SN \citep {2016MNRAS.456..853P}. The transition to redder colours for SNe~2019uo and 2010al places their behavior between SNe~Ib and most extreme SNe~Ibn. SN~2006jc \citep{2007Natur.447..829P} and OGLE-2012-SN-006 \citep{2015MNRAS.449.1941P} show redder colours post 50 days. 

\section{Spectral evolution}
\label{4}
\begin{figure}
	\begin{center}
		\includegraphics[width=\columnwidth]{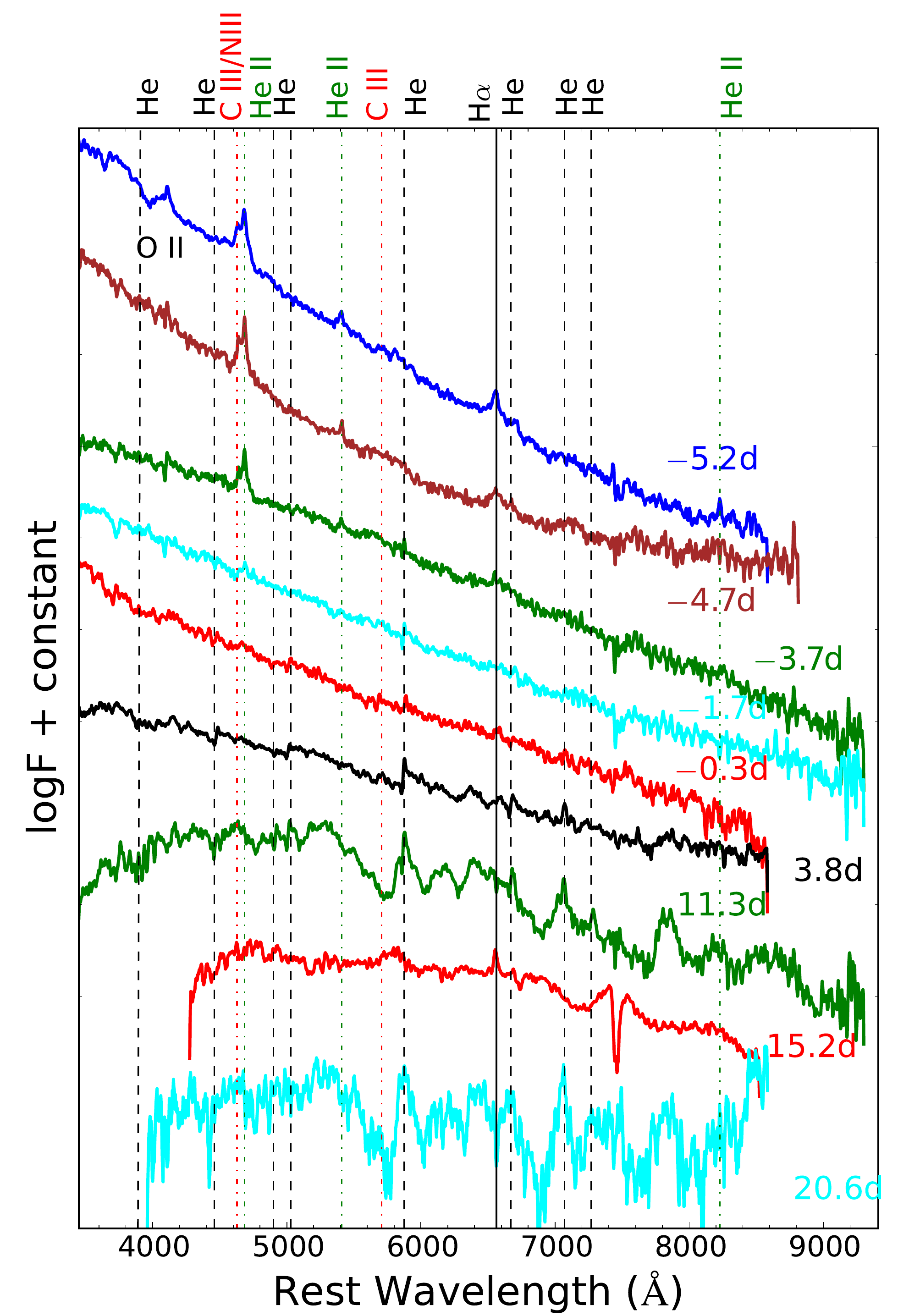}
	\end{center}
	\caption{Spectral evolution of SN~2019uo from $-5.2$~days to 20~days post r$_{max}$. Prominent He features are seen in the early spectra. Flash ionization signatures of \ion{He}{2}, \ion{C}{3} and \ion{N}{3} are also seen.}
	\label{fig:spec_early}
\end{figure}

\begin{figure}
	\begin{center}
		\includegraphics[width=\columnwidth]{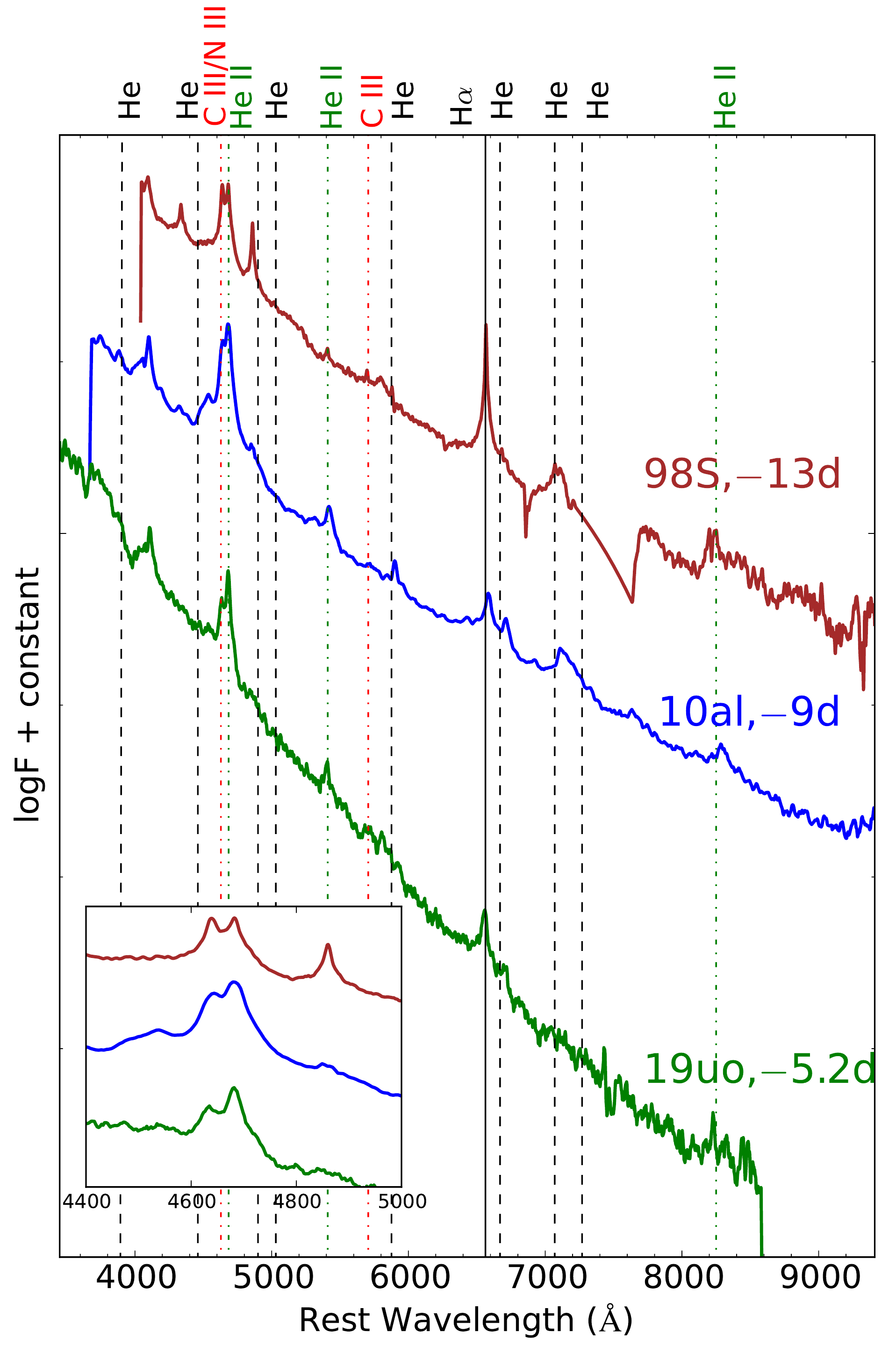}
	\end{center}
	\caption{Spectral comparison of SN~2019uo at $-5.2$~days to SNe~1998S \citep{2001MNRAS.325..907F} and 2010al \citep{2015MNRAS.449.1954P}. Prominent flash ionization features are marked.}
	\label{fig:comp_early}
\end{figure}

The spectral evolution of SN~2019uo from $-5.2$~days to 20.6~days post maximum is displayed in Figure~\ref{fig:spec_early}. The early spectral sequence shows a unique blue continuum similar to SN~2010al. Blackbody fits to the first three spectra ($-5.2$, $-4.7$, and $-3.7$ days) show that the photospheric temperature varies between 13,000 K and 10,000 K. A very narrow H emission line ($\leq$137~km~s$^{-1}$; unresolved) in the early spectrum of SN~2019uo is most likely due to interstellar gas in the host galaxy. Prominent emission features in the first three spectra ($-5.2$ to $-3.7$~days) of SN~2019uo are seen around $\sim$4660 \AA. The emission components are double-peaked, with the blue component peaking at 4643 \AA\ and the red component peaking at 4682 \AA. The red component at 4682 \AA\ is due to \ion{He}{2} at 4686 \AA, whereas the blue component arises from a blend of \ion{C}{3} 4648 \AA\ and \ion{N}{3} 4640 \AA. Another interesting feature is the possible identification of a doubly ionized \ion{C}{3} feature at 5696 \AA. \cite{2015MNRAS.449.1921P} interpreted these as flash ionization signatures in a He-rich CSM (also see \citealp{2014AAS...22323502G}). Although \ion{C}{3} features were found in PTF12ldy and iPTF15ul \citep{2017ApJ...836..158H}, SN~2010al is the only previous SN~Ibn where flash ionization signatures of \ion{C}{3} and \ion{He}{2}, typical of SNe~II, are both seen. \cite{2010ATel.2491....1C} and \cite{2010CBET.2223....1S} identified such lines to be originating from a WR wind, previously noted in SNe~IIn (e.g., SN~1998S; \citet{2001MNRAS.325..907F} and SN~2008fq; \citet{2013A&A...555A..10T}). We also identify a \ion{He}{2} 5411 \AA\ feature with a velocity of 1483~km~s$^{-1}$ at $-5.2$~days. In the spectrum at $-5.2$~days, we see a deep absorption feature at $\sim$4000 \AA\ and a small dip around 8200 \AA, which is likely due to the presence of \ion{O}{2} and \ion{He}{2} features, respectively.

Figure~\ref{fig:comp_early} shows the spectra of SNe~1998S (type~IIn) and 2010al (type~Ibn) in comparison with SN~2019uo. These two SNe have previously shown flash ionization signatures. While the spectrum of SN~2010al shows \ion{C}{3} features around 4650 \AA\ only, SN~2019uo shows \ion{C}{3} features around 4650 \AA\ and at 5696 \AA. The inset in Figure~\ref{fig:comp_early} highlights these features.

\begin{figure}
	\begin{center}
		\includegraphics[width=\columnwidth]{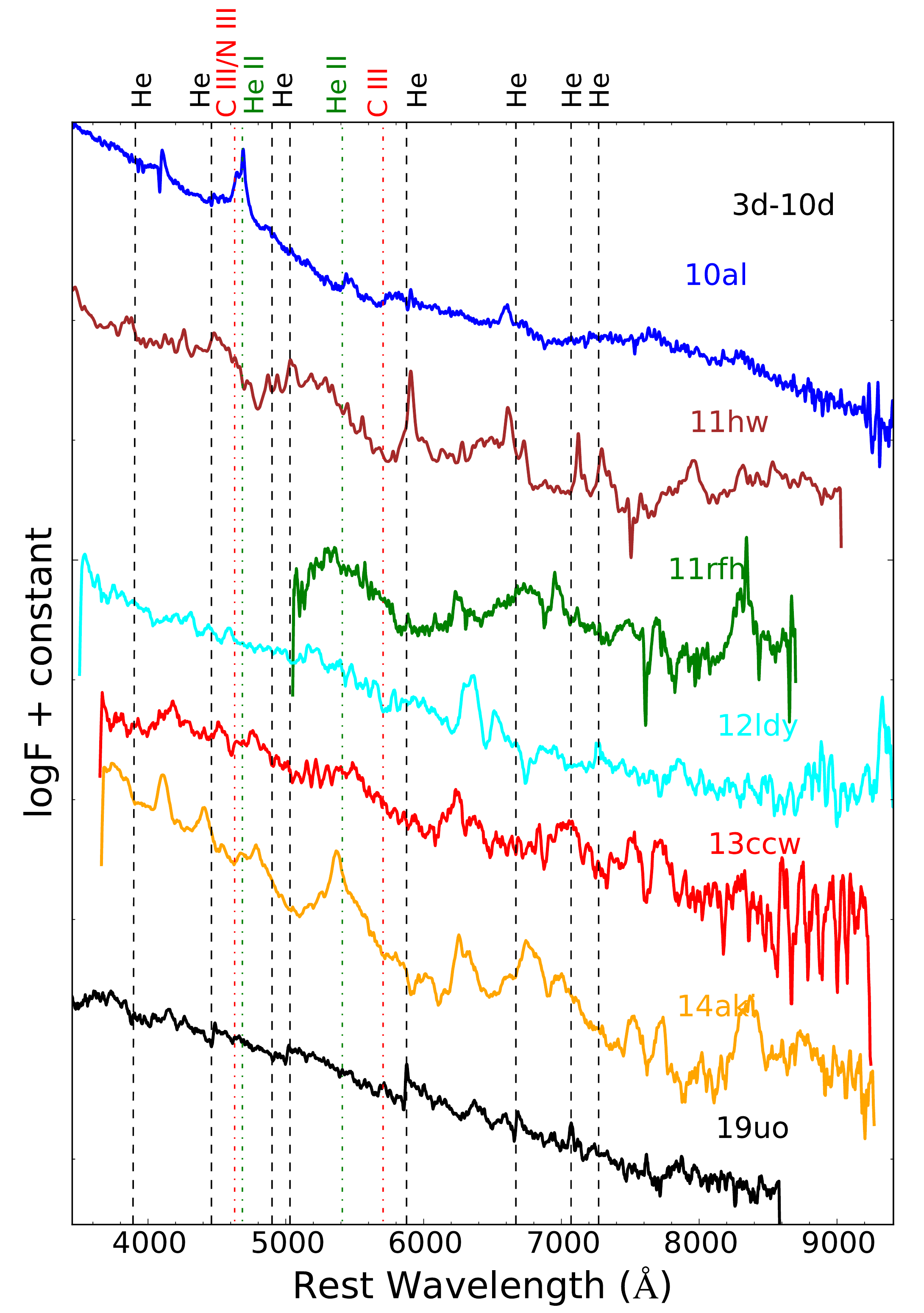}
	\end{center}
	\caption{Comparison of the spectrum of SN~2019uo to other SNe~Ibn. SN~2019uo and SN~2010al show distinct narrow P-cygni \ion{He}{1} spectroscopic features. The data for this are taken from --- SNe~2010al \citep{2015MNRAS.449.1921P}, 2011hw \citep{2015MNRAS.449.1921P}, PTF11rfh \citep{2017ApJ...836..158H},  PTF12ldy \citep{2017ApJ...836..158H}, LSQ13ccw \citep{2015MNRAS.449.1954P} and iPTF14aki \citep{2017ApJ...836..158H}}
	\label{fig:comp_mid}
\end{figure}

\begin{figure*}
	\begin{center}
		\includegraphics[width=1.0\textwidth]{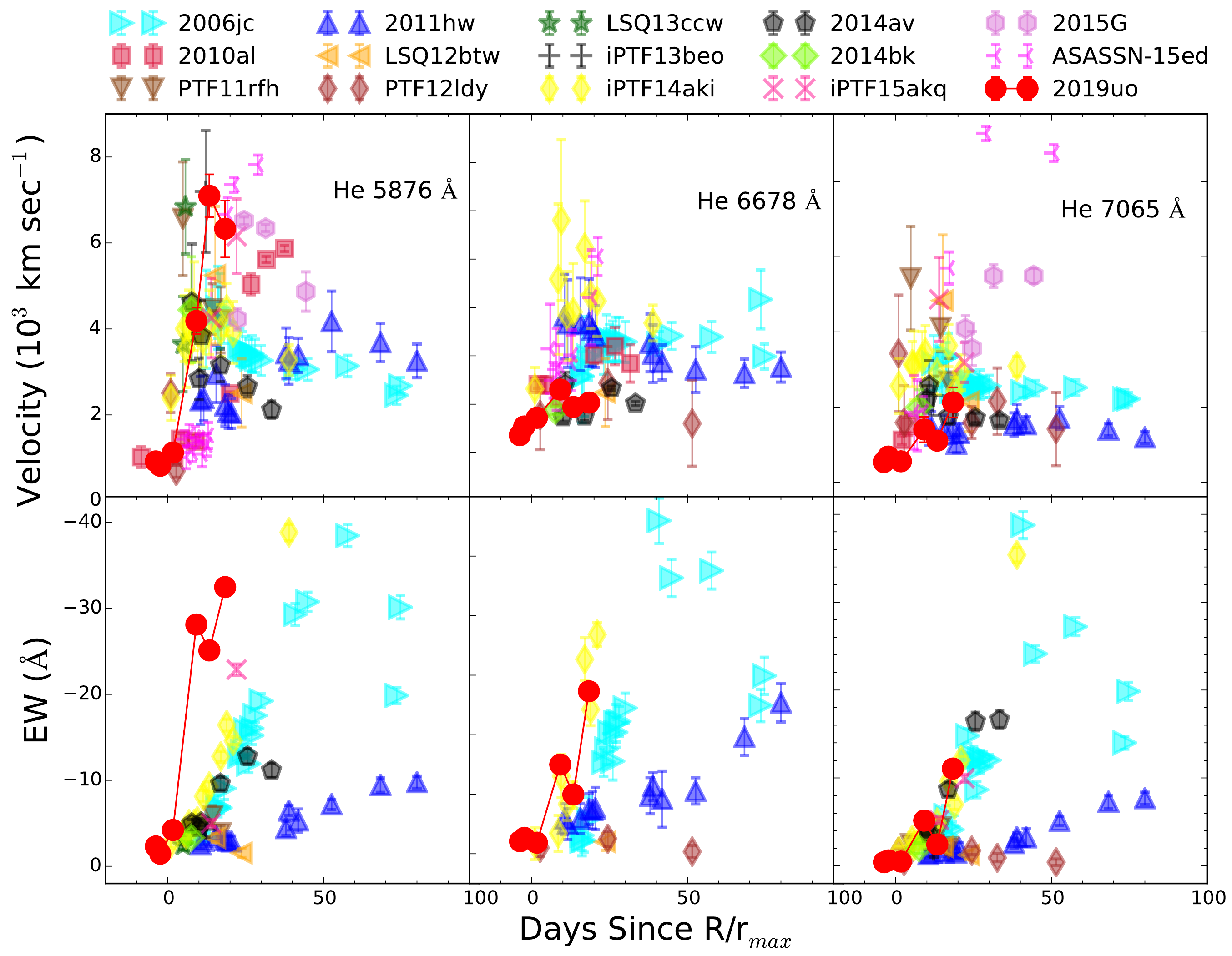}
	\end{center}
	\caption{Evolution of line velocities and equivalent widths of \ion{He}{1} emission lines is shown in top and bottom panels, respectively. The data for this are taken from --- SNe~2006jc \citep{2007ApJ...657L.105F,2008MNRAS.389..113P}, 2010al \citep{2015MNRAS.449.1921P}, 2011hw \citep{2015MNRAS.449.1921P}, PTF11rfh \citep{2017ApJ...836..158H}, LSQ12btw \citep{2015MNRAS.449.1954P}, OGLE12-006 \citep{2015MNRAS.449.1941P}, PTF12ldy \citep{2017ApJ...836..158H}, iPTF13beo \citep{2014MNRAS.443..671G}, LSQ13ccw \citep{2015MNRAS.449.1954P}, iPTF14aki \citep{2017ApJ...836..158H}, 2014av \citep{2016MNRAS.456..853P}, 2014bk \citep{2016MNRAS.456..853P}, iPTF15akq  \citep{2017ApJ...836..158H}, ASASSN-15ed \citep{2015MNRAS.453.3649P}, and 2015G \citep{2017ApJ...836..158H}.}
	\label{fig:vel_eq}
\end{figure*}

As the SN evolves further (3.8~days), the narrow \ion{He}{1} P~Cygni feature is superimposed on a broader base (the continuum is not flat). The flash ionization spectral features vanish completely during this epoch. From 11--21~days, features of \ion{Ca}{2}, \ion{Si}{2}, and \ion{Na}{1}D also start developing (see Figure~\ref{fig:spec_early}). Figure~\ref{fig:comp_mid} shows the comparison of SN~2019uo with a group of SNe~Ibn between 3 -- 10 days after peak. The \ion{He}{1} 5876 \AA\ feature of SN~2019uo is similar to that identified in SN~2010al. However, the \ion{He}{1} P~Cygni feature of SN~2019uo is narrower, and is superimposed over a broader emission line. On the other hand, the \ion{He}{1} P-Cygni profile in SN~2010al is over a flat continuum. Flash ionization signatures in SN~2010al are still visible at this phase, but these features have vanished in SN~2019uo. The line evolution of SN~2019uo shows that it belongs to the ``P~Cygni" subclass (following the interpretation of \citealt{2017ApJ...836..158H}). The P~Cygni \ion{He}{1} features are narrow but gradually broaden with time. The physical explanation behind the origin of the ``P~Cygni" subclass could be a shell of He around the progenitor star surrounded by a dense CSM. As the optically thick shell is lit by the explosion, the narrow P~Cygni features transition to broader emission as the shell is swept up by the SN ejecta. The viewing angle dependence could also affect this scenario; if the CSM is asymmetric and we have a He~rich torus, then P~Cygni features would only be visible if the system is viewed edge-on, while emission features can be seen only if it is viewed face-on. However, this scenario was questioned by \cite{2019arXiv191006016K} which suggested that \ion{He}{1} line fluxes are largely dependent on density, temperature and optical depths. \cite{2019arXiv191006016K} suggest that dominance of emission at late phases is not because of being optically thin, but because they lack other lines to branch into it. He ionisation and recombination are mostly caused by UV and X-ray, occurring at shock boundary, deep in interacting regions. Even though most of the emission and the electron scattering are produced by the ionised region outside the shock, P-cygni features usually originate from optical depths $\leq$ 1. X-rays penetrating further into the P-cygni producing regions will fill in the absorption and lead to emission features. Thus, this provides an alternative scenario to the transitioning of P-cygni to emission features of \ion{He}{1} lines for type Ibn SNe.

The measured the expansion velocities and equivalent widths (EWs) of three neutral He lines (5876, 6678, and 7065 \AA), wherever visible. We fit the emission lines of \ion{He}{1} using a Gaussian on a linear continuum. The EW is estimated through the integral of the flux normalized to the local continuum. We do not measure the EW of the P-cygni lines. The velocities reported are estimated from the absorption minima of P-cygni profiles. Figure~\ref{fig:vel_eq} shows the evolution of velocity and EW for a sample of SNe~Ibn taken from \cite{2017ApJ...836..158H} with time. We see that both the line velocities and EW of the He lines gradually increase with time and the velocity estimates of SN~2019uo lie in the lower range of SNe~Ibn. However, SN~2019uo shows a faster evolution in line velocities, reaching broader emission profiles as seen in the P-cygni subclass \citep{2017ApJ...836..158H} while the emission subclass shows very little velocity evolution.

\begin{figure}
	\begin{center}
		\includegraphics[width=\columnwidth]{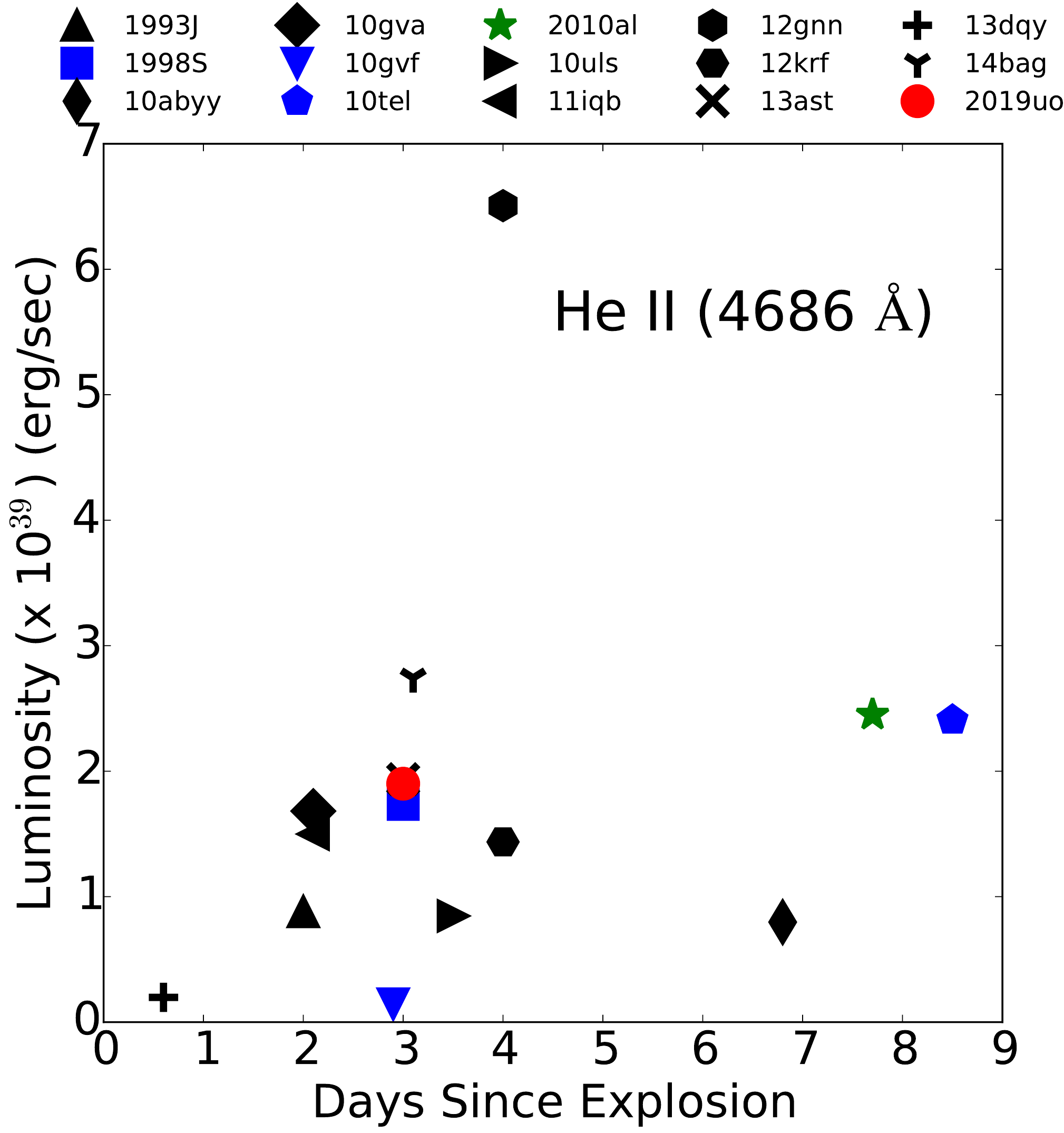}
	\end{center}
	\caption{\ion{He}{2} luminosity of a sample of SNe~II, IIn, and Ibn \citep{2016ApJ...818....3K} with flash ionization signatures. Blue symbols: type IIn, Black symbols: type II (IIb,IIP and IIL),  Red symbols: type Ibn.}
	\label{fig:flash_ionization.}
\end{figure}
To ascertain the origin of the SNe~Ibn, we collected a sample of 12 SNe~II (including SNe~IIb and IIP, IIn) and Ibn from \cite{2016ApJ...818....3K} that showed signatures of flash ionization within 10 days of explosion. Since the H lines are usually contaminated by the host galaxy, we selected the relatively unblended \ion{He}{2} 4686 \AA\ line. Since the \ion{He}{2} lines are much narrower than lines from the SN ejecta, they can serve as a good tool for probing the flash-ionized CSM. When measuring the luminosities, we removed the continuum by fitting a linear function. Figure~\ref{fig:flash_ionization.} shows that the typical luminosity of the \ion{He}{2} line for SN~2019uo is similar to the type~IIn SNe 1998S and PTF13ast. 

\section{Modeling the bolometric light curve of SN~2019uo}
\label{3.2} 
To construct the bolometric light curve of SN~2019uo, the measured flux values were corrected for distance and reddening as given in Section~\ref{1}. Spectral energy distributions (SEDs) were constructed accounting for the flux coverage between UV to IR bands using the {\sl SuperBol} \citep{2018RNAAS...2..230N} code. The lack of UV and NIR data was supplemented by extrapolating the SEDs using the blackbody approximation and direct integration method as described in \cite{2017PASP..129d4202L}. A linear extrapolation was performed in UV regime at late times. The estimated peak bolometric luminosity of SN~2019uo is $8.9 \times 10^{42}$~erg~s$^{-1}$. We used different models to fit the bolometric light curve at a fixed optical opacity of 0.1~cm$^{2}$~g$^{-1}$. A Markov Chain Monte Carlo (MCMC) technique was used to obtain the best-fit parameters. 
	\begin{figure}
		\begin{center}
			\includegraphics[width=0.5\textwidth,angle=0]{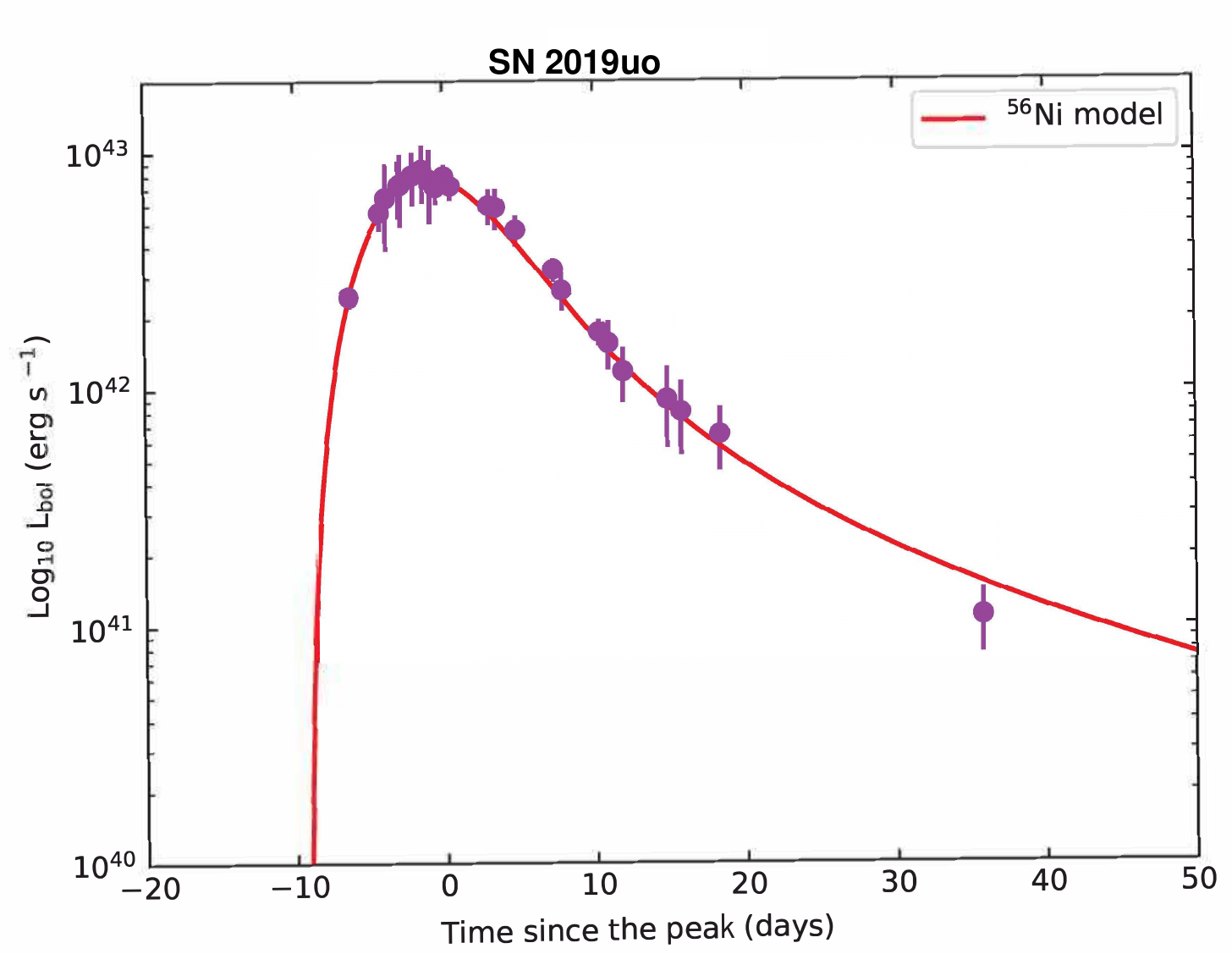}
		\end{center}
		\caption{Best-fit light curves of SN~2019uo using a $^{56}$Ni model.}
		\label{fig:ni+mag}
	\end{figure} 
	
	\begin{figure*}
		\begin{center}
				\includegraphics[width=\textwidth,angle=0]{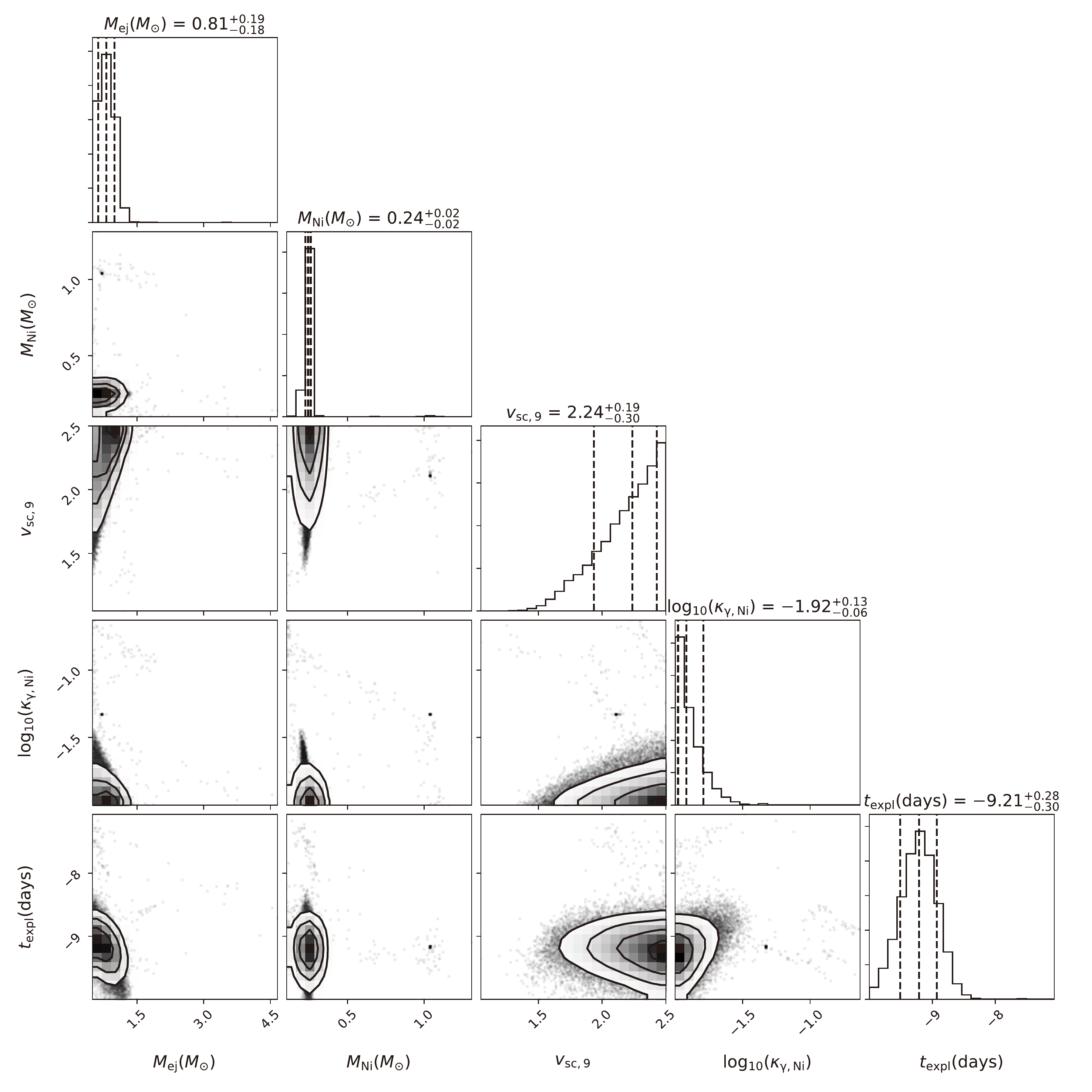}
		\end{center}
		\caption{The corner plot of the $^{56}$Ni model displaying covariance of estimated parameters.}
		\label{fig:corner_ni}
	\end{figure*} 

{\bf $^{56}$Ni model:} Assuming that the peak bolometric luminosity is powered by the decay of $^{56}$Ni to $^{56}$Co, we fit the bolometric light curve using $^{56}$Ni model \citep{1982ApJ...253..785A,1980ApJ...237..541A}. The parameters of the $^{56}$Ni model are the ejecta mass $M_\mathrm{ej}$, the initial scale velocity of the ejecta $v_{\mathrm{sc}0}$, the $^{56}$Ni mass $M_\mathrm{Ni}$, the gamma-ray opacity of $^{56}$Ni decay photons $\kappa_{\gamma,\mathrm{Ni}}$ and explosion time $t_\mathrm{expl}$. The initial kinetic energy of the ejecta is neutrino-driven and is considered to be $E_\mathrm{k} = 0.3 M_\mathrm{ej} v_{\mathrm{sc}0}^2$. The best-fit parameters are tabulated in Table~\ref{tab:para1} and the best-fit model is displayed Fig.~\ref{fig:ni+mag}. The corner plot showing the covariance of the estimated parameters are represented in Fig.~\ref{fig:corner_ni}. We note that the $^{56}$Ni mass obtained from the powering mechanism of \cite{1982ApJ...253..785A} are in concordance with the values quoted for several stripped envelope SNe \citep{2016MNRAS.457..328L,2016MNRAS.458.2973P,2019MNRAS.485.1559P}. Although the $^{56}$Ni mass inferred from the model is $\sim$ 0.24 M$_\odot$ which is comparable to that of normal CCSNe, the opacity for the gamma ray $\kappa_{\gamma,\mathrm{Ni}}$ emitted from the cascade decay of $^{56}$Ni is 0.01 cm$^2$g$^{-1}$, which is significantly smaller than the canomical lower limit which is 0.025-0.027 cm$^2$g$^{-1}$. Therefore, the $^{56}$Ni model is not a good model in explaining the light curve of SN~2019uo and other models must be employed.


\setlength{\tabcolsep}{1mm}
\begin{table}
	\caption{Parameters of the $^{56}$Ni model. The uncertainties are $1\sigma$.}
	\label{tab:para1}
	\begin{center}
		{\scriptsize
			\begin{tabular}{ccccccccccc}
				\toprule
				&	 $M_{\mathrm{ej}}$ 	&	 $M_{\mathrm{Ni}}$ 	&	 $v_{\mathrm{sc}0}$ 	&	 $\kappa_{\gamma,\mathrm{Ni}}$  	& $t_\mathrm{expl}^{\star}$	&	$\chi^2/\mathrm{dof}$ \\
				&	 (M$_{\odot}$) 	&	 (M$_{\odot}$) 	&	 ($10^9$cm s$^{-1}$) 	&	 (cm$^2$ g$^{-1}$) 	&		 (days)	&	\\
                \midrule
				&	$0.81_{-0.18}^{+0.19}$	&	$0.24_{-0.02}^{+0.02}$	&	$2.24_{-0.30}^{+0.19}$	&	$0.01_{-0.00}^{+0.00}$	&	$-9.21_{-0.30}^{+0.28}$	&	$6.67/19$\\
				\bottomrule
			\end{tabular}}
		\end{center}
		\par
		{$\star$ The value of $t_\mathrm{expl}$ is with respect to $r_{max}$. \newline}
	\end{table}

	\setlength{\tabcolsep}{1mm}
	\begin{table*}
		\caption{Parameters of the CSI model and the CSI plus $^{56}$Ni model. The uncertainties are 1$\sigma$.}
		\label{tab:para2}
		\begin{center}
			{\scriptsize
				\begin{tabular}{cccccccccccccc}
					\toprule
					&	$s$	&	 $E_{\mathrm{SN}}$	&	 $M_{\mathrm{ej}}$ 	&	 $M_{\mathrm{Ni}}$ 	&	 $M_{\mathrm{CSM}}$ 	&	 $\rho_{\mathrm{CSM,in}}$ 	&	 $R_{\mathrm{CSM,in}}$ 	&	$\epsilon$ 	&	$x_{\mathrm{0}}$	&	 $\kappa_{\gamma,\mathrm{Ni}}$ 	&	 $t_\mathrm{expl}^{\star}$	&	 $\chi^2/\mathrm{dof}$ \\\\
					&	 	&	 ($10^{51}$~erg) 	&	 ($M_\odot$) 	&	 ($M_\odot$) 	&	 ($M_\odot$) 	&	 ($10^{-12}$g cm$^{-3}$) 	&	 ($10^{14}$cm) 	&	 	&		&	 (cm$^2$ g$^{-1}$)  	&	 (days)	&	\\
					\midrule
					CSI	&	2	&	$0.87_{-0.04}^{+0.06}$	&	$8.83_{-0.99}^{+0.71}$	&	\nodata	&	$0.40_{-0.03}^{+0.04}$	&	$3.34_{-1.72}^{+3.33}$	&	$1.76_{-0.55}^{+0.91}$	&	$0.11_{-0.01}^{+0.01}$	&	$0.35_{-0.10}^{+0.08}$	&	\nodata	&	$-7.24_{-0.08}^{+0.09}$	&	$3.95/16$\\\\
					CSI	&	0	&	$0.40_{-0.14}^{+0.31}$	&	$13.51_{-5.19}^{+3.91}$	&	\nodata	&	$1.28_{-0.44}^{+0.41}$	&	$0.15_{-0.04}^{+0.12}$	&	$19.05_{-7.98}^{+6.66}$	&	$0.51_{-0.25}^{+0.29}$	&	$0.67_{-0.22}^{+0.20}$	&	\nodata	&	$-7.89_{-0.07}^{+0.07}$	&	$13.44/16$\\\\
					CSI+$^{56}$Ni 	&	2	&	$1.67_{-0.23}^{+0.18}$	&	$15.99_{-2.98}^{+2.25}$	&	$0.01_{-0.002}^{+0.003}$	&	$0.41_{-0.07}^{+0.08}$	&	$20.96_{-4.83}^{+4.73}$	&	$8.04_{-1.39}^{+1.49}$	&	$0.64_{-0.12}^{+0.14}$	&	$0.51_{-0.19}^{+0.25}$	&	$0.95_{-0.88}^{+10.15}$	&	$-6.42_{-0.00}^{+0.00}$	&	$2.79/14$\\\\
					CSI+$^{56}$Ni 	&	0	&	$1.78_{-0.19}^{+0.13}$	&	$16.30_{-2.72}^{+2.09}$	&	$0.01_{-0.002}^{+0.003}$	&	$0.73_{-0.11}^{+0.12}$	&	$25.05_{-3.58}^{+2.75}$	&	$14.16_{-2.00}^{+1.85}$	&	$0.71_{-0.12}^{+0.12}$	&	$0.43_{-0.14}^{+0.22}$	&	$0.90_{-0.82}^{+10.40}$	&	$-6.40_{-0.00}^{+0.00}$	&	$3.17/14$\\
					\bottomrule
				\end{tabular}}
			\end{center}
			\par
			{$\star$ The value of $t_\mathrm{expl}$ is with respect to $r_{max}$. \newline}
		\end{table*}
		
		\begin{figure*}
			\begin{center}
				\includegraphics[width=0.45\textwidth,angle=0]{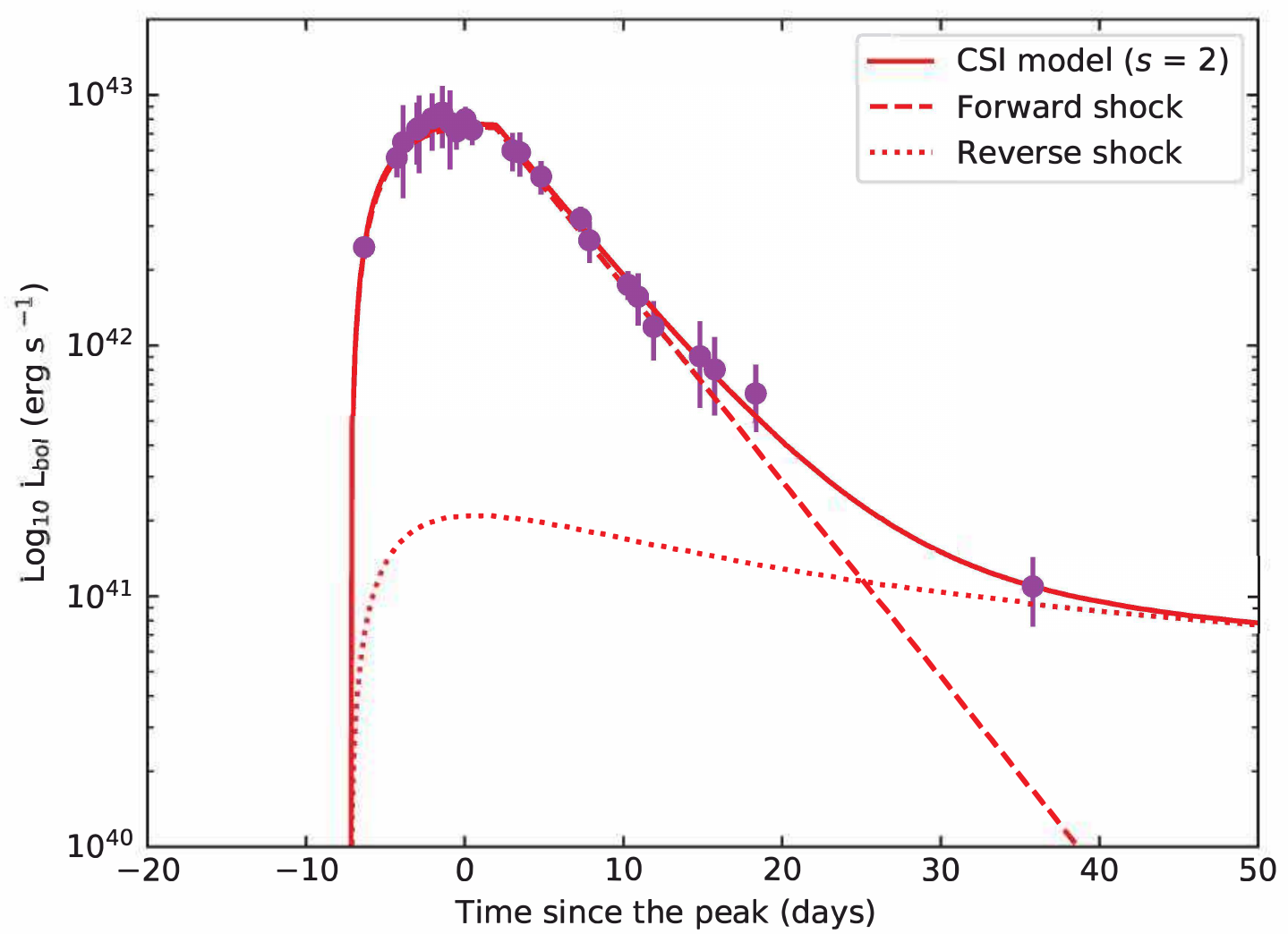}
				\includegraphics[width=0.45\textwidth,angle=0]{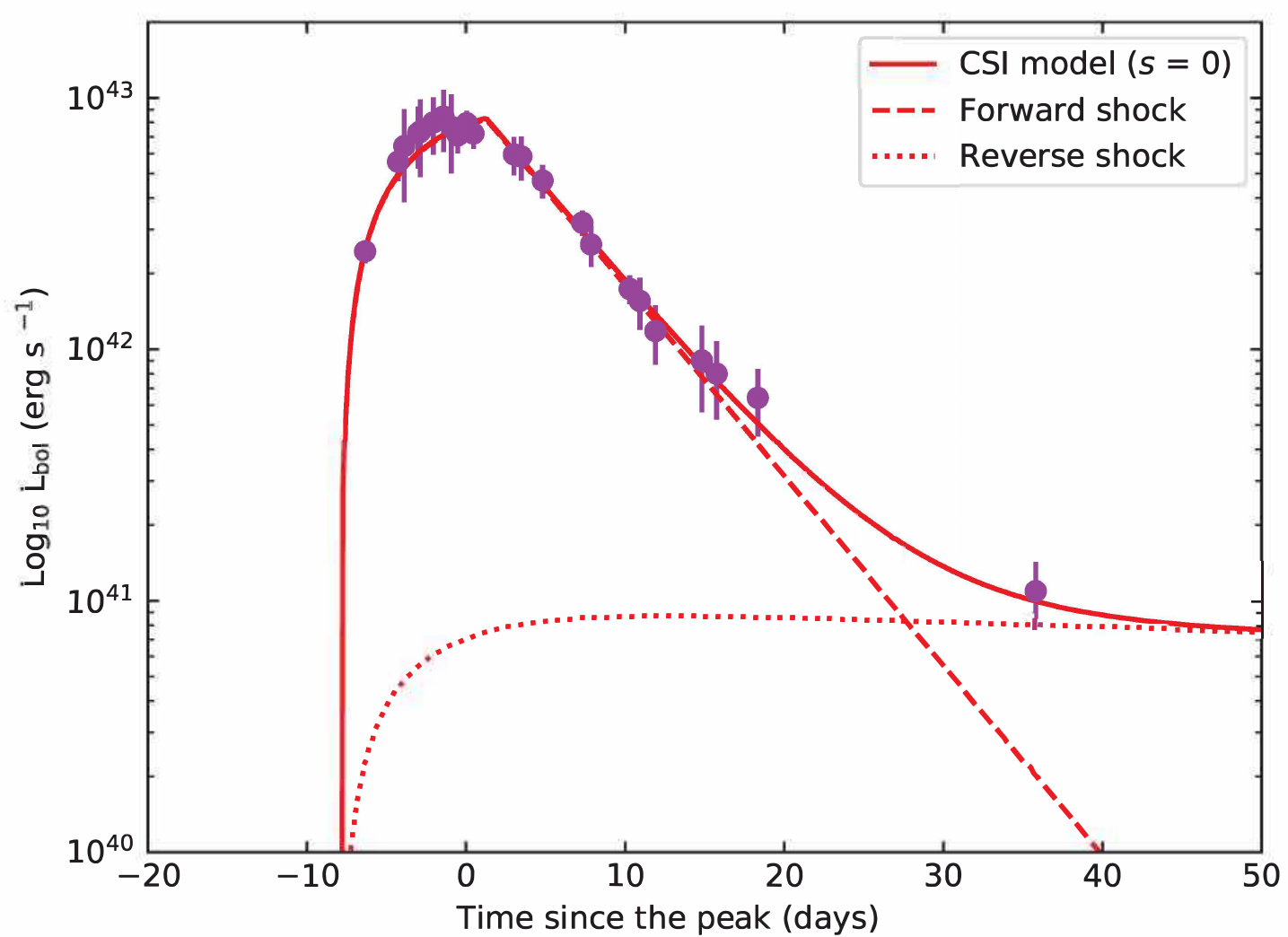}
				\includegraphics[width=0.45\textwidth,angle=0]{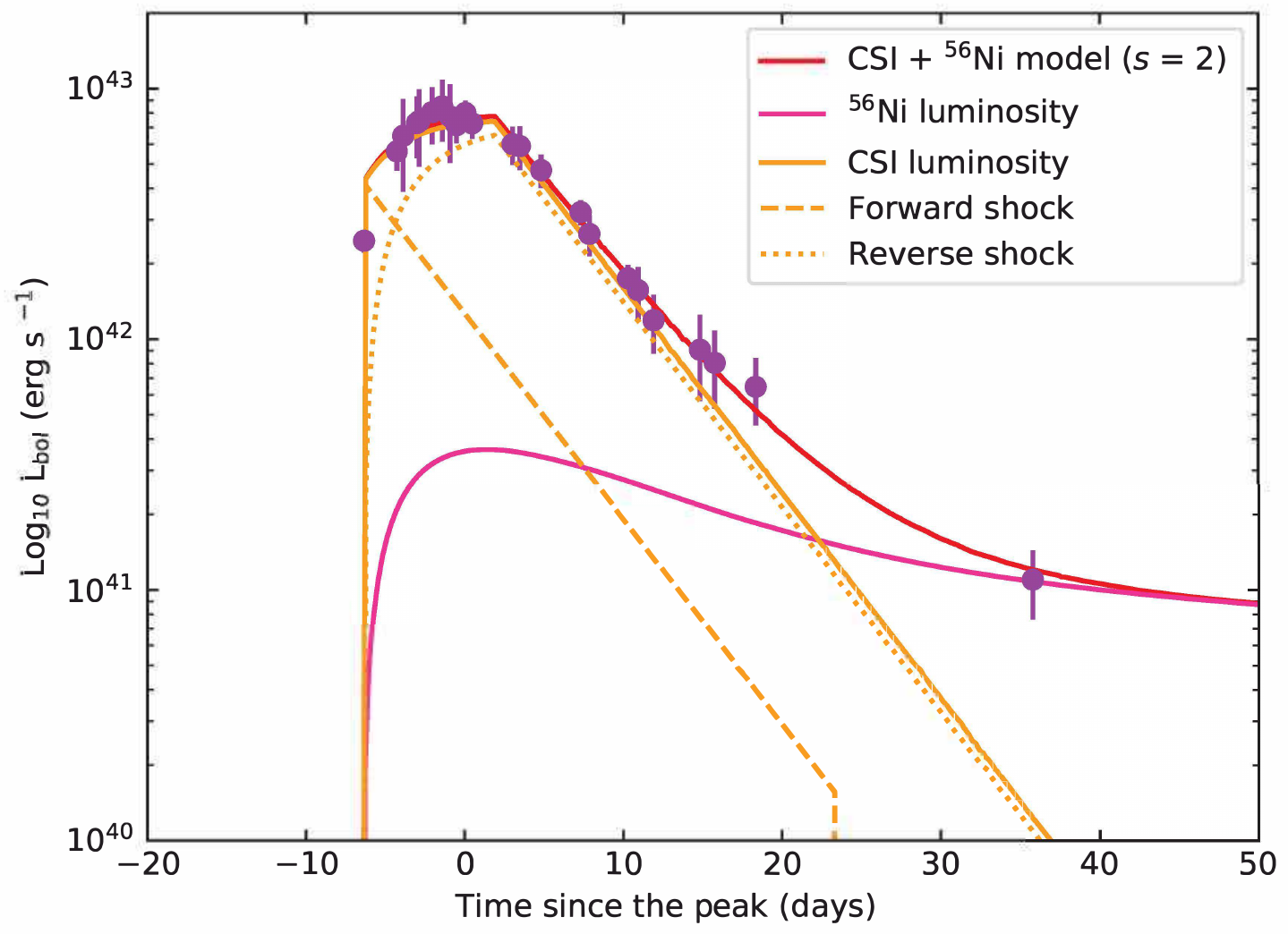}
				\includegraphics[width=0.45\textwidth,angle=0]{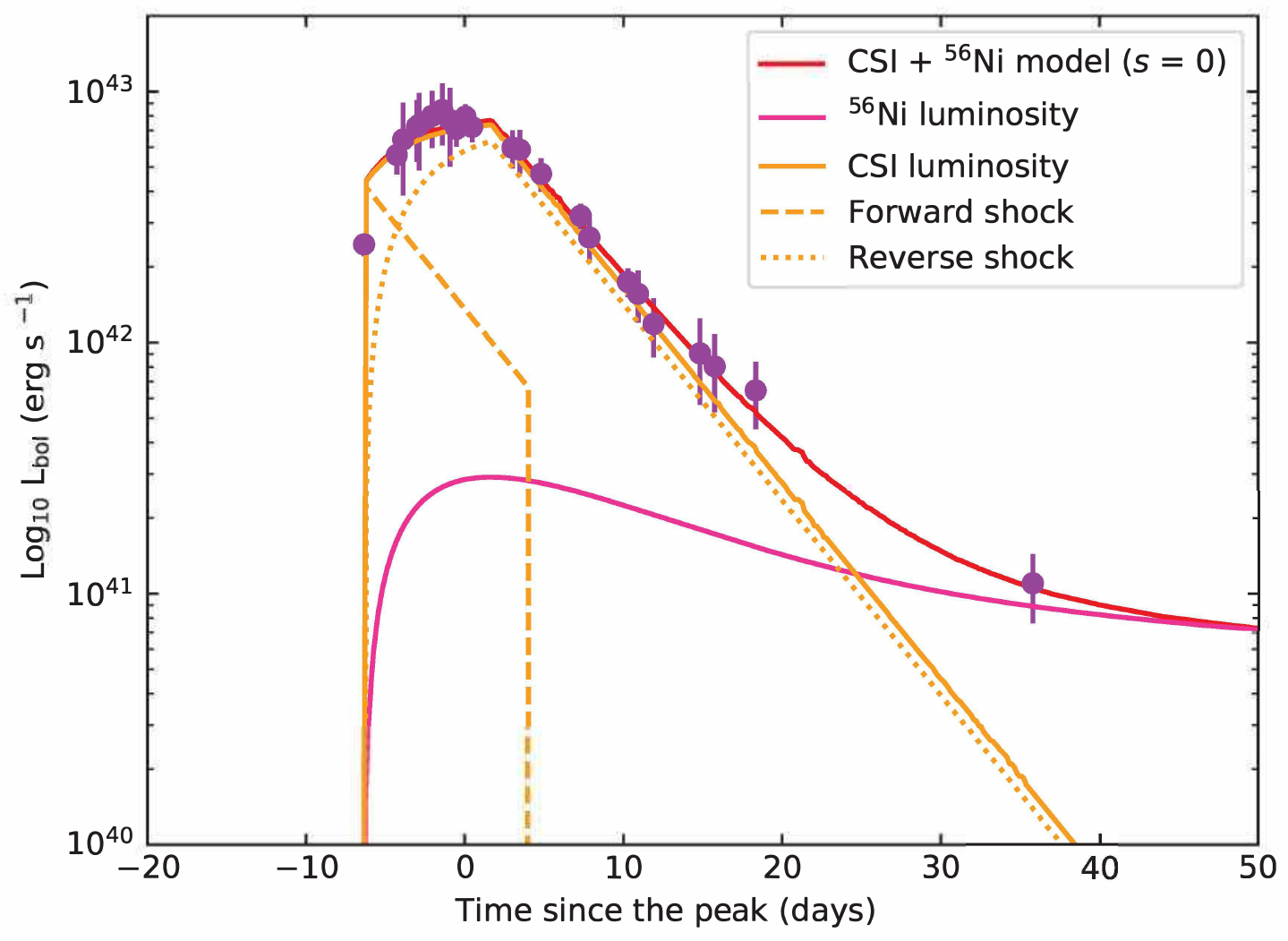}
			\end{center}
			\caption{Best-fit light curves of SN~2019uo fitted with a CSI model and a combination of $^{56}$Ni and CSI. The forward shocks, reverse shocks, and $^{56}$Ni models are plotted with different lines. }
			\label{fig:ni+CSI}
		\end{figure*}
		
		\begin{figure*}
			\begin{center}
				\includegraphics[width=\textwidth,angle=0]{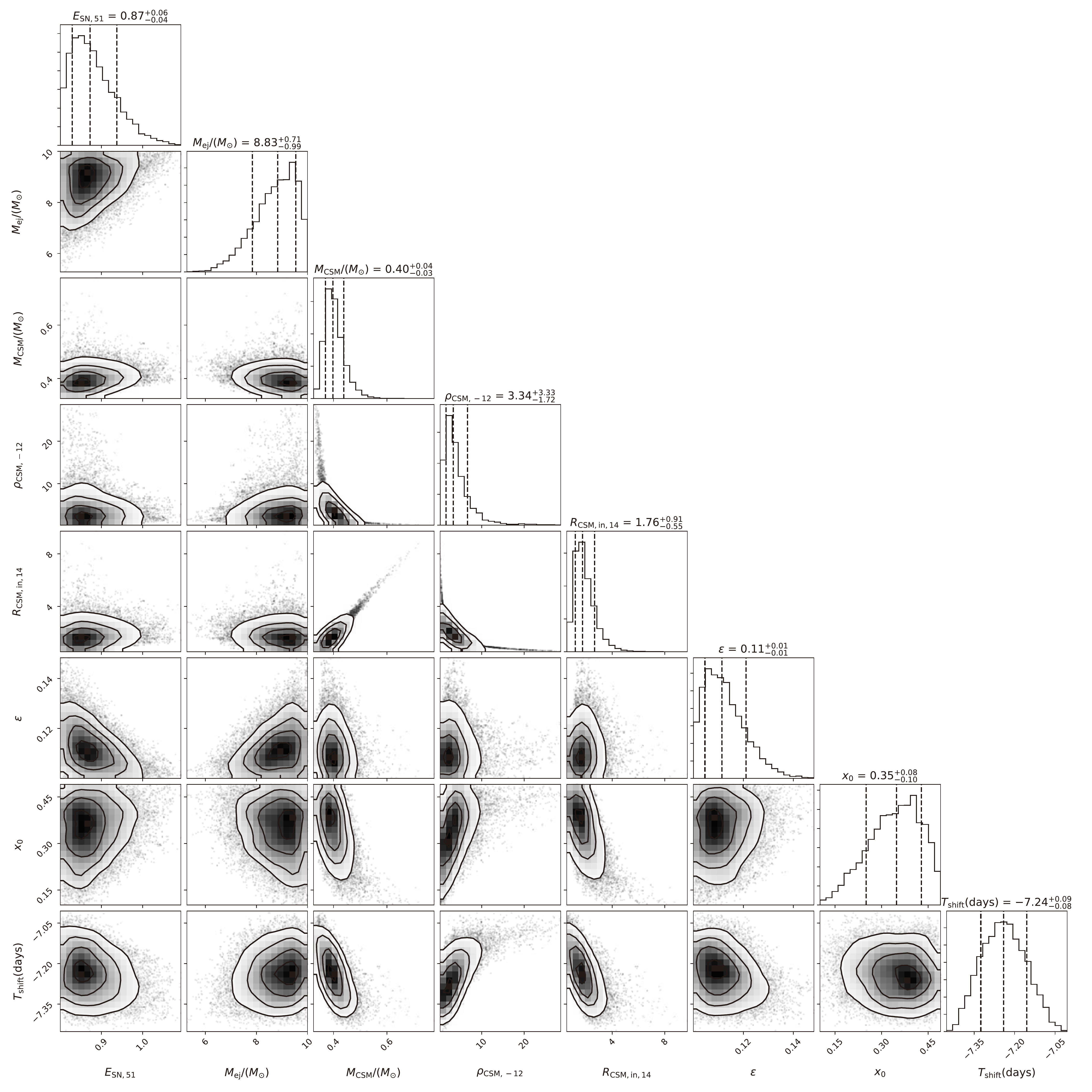} \\
			\end{center}
			\caption{The corner plot of the CSI wind model displaying covariance of estimated parameters.}
			\label{fig:corner_windCSI}
		\end{figure*}
				\begin{figure*}
			\begin{center}
				\includegraphics[width=\textwidth,angle=0]{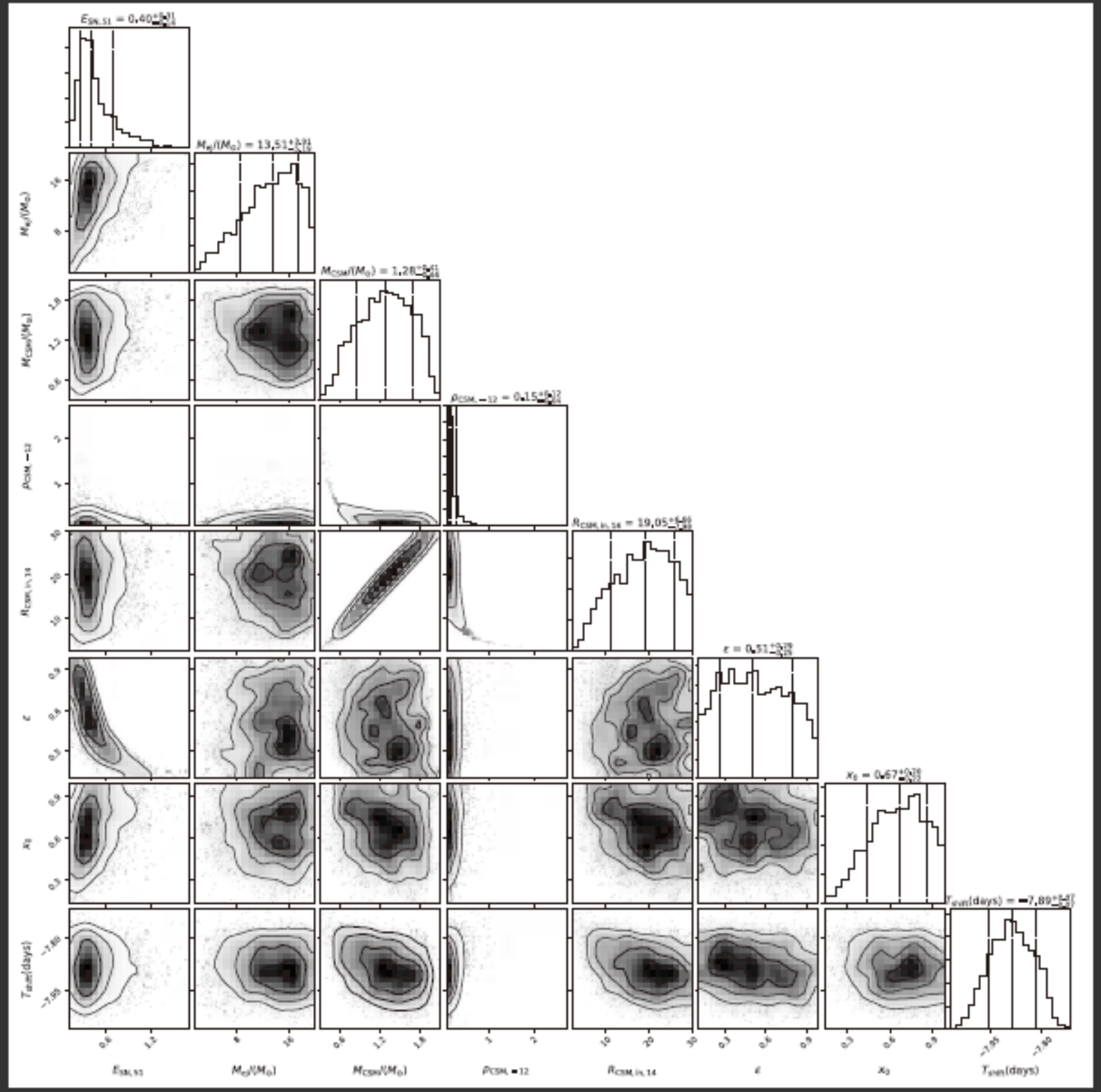} \\
			\end{center}
			\caption{The corner plot of the CSI shell model displaying covariance of estimated parameters.}
			\label{fig:corner_shellCSI}
		\end{figure*}
		
		\begin{figure*}
			\begin{center}
				\includegraphics[width=\textwidth,angle=0]{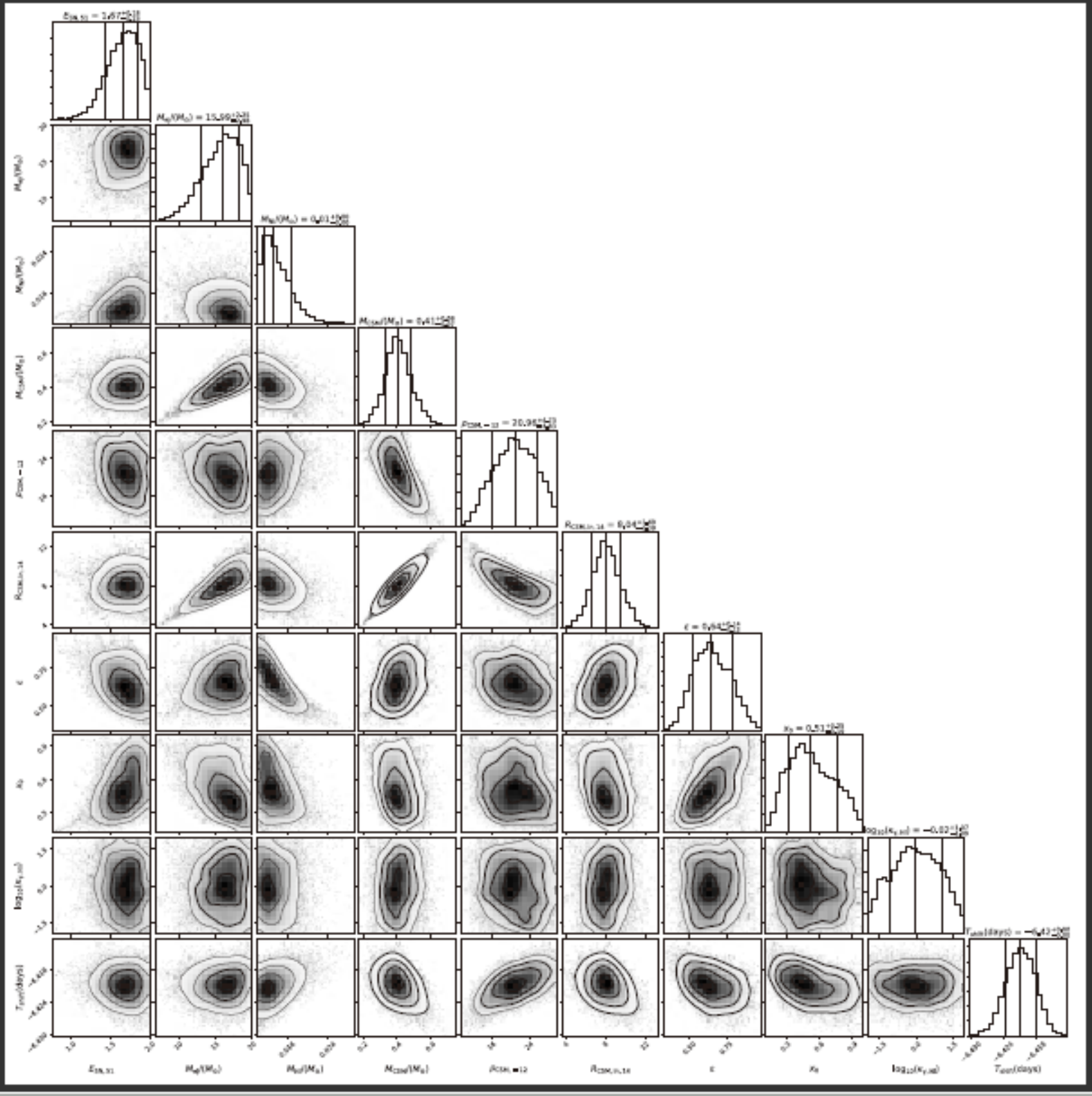} \\
			\end{center}
			\caption{The corner plot of the $^{56}$Ni + CSI wind model displaying covariance of estimated parameters.}
			\label{fig:corner_ni+CSIwind}
		\end{figure*}
				\begin{figure*}
			\begin{center}
				\includegraphics[width=\textwidth,angle=0]{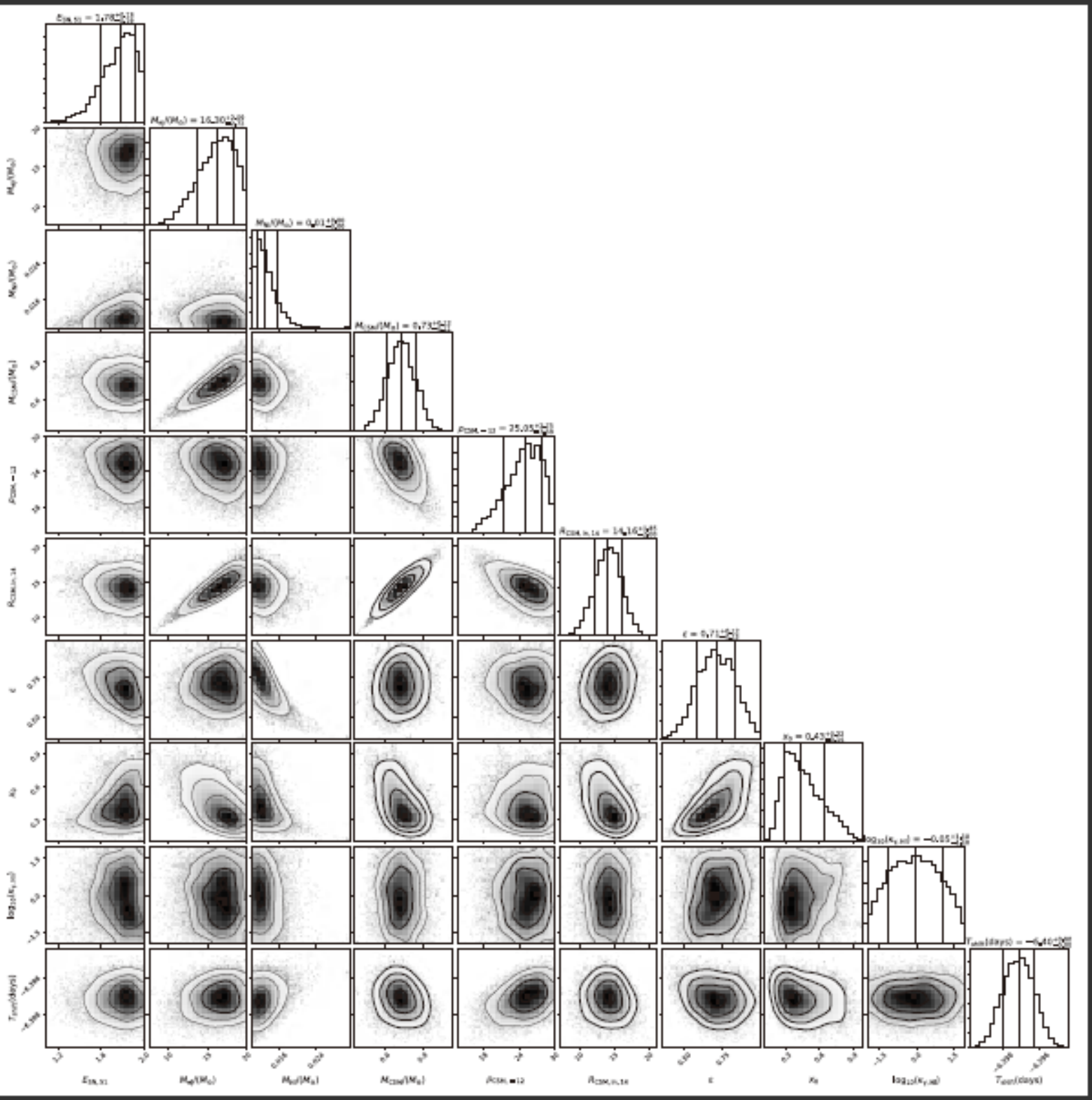} \\
			\end{center}
			\caption{The corner plot of the $^{56}$Ni + CSI shell model displaying covariance of estimated parameters.}
			\label{fig:corner_ni+CSIshell}
		\end{figure*}
		
		{\bf The CSI model and the $^{56}$Ni + CSI model:} The narrow He emission lines appearing in the spectra of SN~2019uo indicate a potential source of circumstellar interaction (CSI) with a nearby He-rich shell. Thus, the nearby He-rich wind or shell surrounding the progenitor could be the essential powering source of the bolometric light curve of SN~2019uo. We take into account the ejecta-CSM interaction model (i.e., the CSI model) \citep{1982ApJ...258..790C,1994ApJ...420..268C,1994MNRAS.268..173C,2012ApJ...757..178G,2018ApJ...856...59L} and the $^{56}$Ni + CSI model \citep{2012ApJ...746..121C}. To fit the bolometric light curve of SN~2019uo, we adopt the formulation given in \cite{2019arXiv190512623W}.
		
		The ejecta can be broadly distinguished into two zones, the inner part ($\rho_\mathrm{ej} \propto r^{-\delta}$) and the outer part ($\rho_\mathrm{ej} \propto r^{-n}$). The density profile of the CSM can typically be described as a power law where $\rho_\mathrm{CSM} \propto r^{-s}$, where $s=0$ corresponds to shells of the CSM and $s=2$ corresponds to winds. Assuming $\delta=1$ and $n=10$, the adopted parameters of the CSM model are the energy of the SN ($E_\mathrm{SN}$), the mass of the ejecta ($M_\mathrm{ej}$), the mass of the CSM ($M_\mathrm{CSM}$), the density of the innermost part of the CSM $\rho_\mathrm{CSM,in}$, the radius of the innermost part of the ejecta $R_\mathrm{CSM,in}$, the efficiency factor which converts kinetic energy to radiation ($\epsilon$), the dimensionless $x_0$ parameter\footnote{$x \equiv \frac{r(t)}{R(t)}$, where $x \leq x_0$ and $x \geq x_0$ are inner and outer parts of the ejecta.}, and t$_{expl}$. Two additional parameters are used in the $^{56}$Ni + CSI model, $M_\mathrm{Ni}$ and $\kappa_{\gamma,\mathrm{Ni}}$. The best-fit parameters of the model are tabulated in Table~\ref{tab:para2} and the best-fit models are displayed in Fig.~\ref{fig:ni+CSI}. The corner plots describing covariance of the parameters are shown in Fig.~\ref{fig:corner_windCSI}, Fig.~\ref{fig:corner_shellCSI}, Fig.~\ref{fig:corner_ni+CSIwind} and Fig.~\ref{fig:corner_ni+CSIshell} respectively. The tabulated values of ejecta masses of the four models are reasonable if the progenitor is a WR star of mass $\sim\!25\ M_\odot$ and the metallicity is nearly solar \citep{2007A&G....48a..35C}. 
		We adopted the $^{56}$Ni, CSI model, and the $^{56}$Ni + CSI models to fit the bolometric light curve of SN~2019uo. The $^{56}$Ni model provides a favourable fit to the light curve, but  this model cannot explain the \ion{He}{1} emission lines present in the spectrum of SN~2019uo. These lines are likely generated because of the CSI. We therefore invoke CSI as the more favourable model to model light curve. For the CSI model, the estimated ejecta masses for $s=0$ and $s=2$ are $8.83^{+0.71}_{-0.99}\ M_\odot$ and $13.51^{+3.91}_{-5.19}\ M_\odot$ respectively. This model, however, did not take into account the role of $^{56}$Ni. Using the combination of both $^{56}$Ni + CSI, the estimated $M_\mathrm{ej}$ for $s=0$ and $s=2$ is $15.99^{+2.25}_{-2.98}\ M_\odot$ and $16.30^{+2.09}_{-2.72}\ M_\odot$, respectively, which are consistent with a WR progenitor scenario. The mass-loss rate is given by $\dot{M} = 4\pi v_\mathrm{w}q$ (where $q = \rho_\mathrm{CSM,in} R_\mathrm{CSM,in}^2$). The velocity of the wind v$_{w}$ = 100-1000 km s$^{-1}$ for WR systems. Considering the wind CSI model (s = 2), we find that the estimated mass-loss rate lies between 0.195-1.95~$M_\odot$~yr$^{-1}$, which is comparable with the values obtained for iPTF13z (0.1--2~$M_\odot$~yr$^{-1}$; \citealp{2017A&A...605A...6N})  and PS15dpn (1--10 $M_\odot$~yr$^{-1}$; \citealp{2019arXiv190512623W}). Using the combination of $^{56}$Ni + CSI model (s = 2), the estimated mass loss rate lies between 25.5-255.4~$M_\odot$~yr$^{-1}$ which is significantly higher than the value obtained for iPTF13z, PS15dpn, and this model can be excluded. Neverthless, the $^{56}$Ni+CSM shell is reasonable.
		
		For the CSM shell and the $^{56}$Ni + CSM shell model, the expelled shell mass prior to explosion are  $\sim$1.3 $M_\odot$ and 0.73 M$_\odot$, respectively. The radius of the inner shell for the $^{56}$Ni + CSI model, as seen from Table~\ref{tab:para2}, is $14 \times 10^{14}$~cm and the typical velocity of WR winds is between 100 and 1000~km~s$^{-1}$ ($10^{7-8}$~cm~s$^{-1}$); so the time at which the shell is expelled prior to explosion is estimated to be between $1.4 \times 10^6$~s and $1.4 \times 10^7$~s, i.e., between 163.8 and 1638.8 days.
		
		\section{Summary}
		\label{5}
		In this paper, we present the photometric and spectral evolution of the type~Ibn SN~2019uo. The typical light curve decay rate of SNe~Ibn is $\sim$0.1~mag d$^{-1}$ in all bands which is in agreement with the decline rates of the SNe Ibn discussed by \cite{2017ApJ...836..158H}. The color evolution of SN~2019uo is similar to SN~2010al and iPTF14aki which places it between SNe~Ib and SNe~Ibn. This is in good agreement with the P~Cygni spectroscopic features that transition from narrow to broad, indicating a He-rich circumstellar shell around the progenitor star along with optically thick CSM \citep{2017ApJ...836..158H}. The absolute magnitude ($M^V_{max}$ = $-$18.30 $\pm$ 0.24 mag) indicates that SN~2019uo lies at the fainter end of the group. We fit the bolometric light curve of SN~2019uo with $^{56}$Ni model. However, the $^{56}$Ni model alone does not take into account the CSM interaction that is evident from the narrow emission lines in the spectra of SN~2019uo. Thus, we also fit the light curves with a CSI model and a $^{56}$Ni + CSI model. The $^{56}$Ni + CSI wind (s=2) model can be excluded since an unrealistic value of mass loss rate (25.5-255.4~$M_\odot$~yr$^{-1}$) is required and the $^{56}$Ni + CSI shell model is reasonable. The combination of $^{56}$Ni +  CSI shell well fits our observed light curve, with ejecta masses consistent with a WR star. The spectroscopic features of SN~2019uo indicate that it is the second SNe~Ibn with flash ionization signatures. Prominent lines of \ion{He}{2}, \ion{C}{3}, and \ion{N}{3} are detected in the spectra, similar to SN~2010al. SN~2019uo shows initial P~Cygni \ion{He}{1} features that broadens after 11 days post-maximum. This can originate from a He-rich shell around progenitor surrounded by dense CSM, or it may be due to viewing angle dependency. This is also validated by the equivalent widths of \ion{He}{1} features. Alternatively, P-cygni spectroscopic features usually originate from optical depths $\leq$ 1. As X-rays penetrate into the P-cygni producing regions absorptions are filled leading to subsequent emission features. The estimated line velocities are lower than the average SN~Ibn, but they show a faster evolution compared to the group of SNe that show prominent emission features from the beginning.
		
		\section*{Acknowledgments}
		We thank the support of the staff of the Xinglong 2.16~m and Lijiang 2.4~m telescope, and observing assistants at the 1.04~m ST, 1.30~m DFOT, and 2.00~m HCT for their support during observations. This work was partially supported by the Open Project Program of the Key Laboratory of Optical Astronomy, National Astronomical Observatories, Chinese Academy of Sciences. Funding for the Lijiang 2.4~m telescope has been provided by Chinese Academy of Sciences and the People's Government of Yunnan Province. We acknowledge Weizmann Interactive Supernova data REPository (WISeREP; \url{http://wiserep.weizmann.ac.il}). DAH acknowledges support from NSF grant AST-1313404.  The work of XW is supported supported by the National Natural Science Foundation of China (NSFC grants 11325313, 11633002, and 11761141001), and the National Program on Key Research and Development Project (grant no.\ 2016YFA0400803). This work makes use of data obtained with the LCO Network. KM and SBP acknowledges BRICS grant DST/IMRCD/BRICS/Pilotcall/ProFCheap/2017(G) for the present work. SBP and KM also acknowledge the DST/JSPS grant, DST/INT/JSPS/P/281/2018. KM acknowledges the support from Department of Science  and Technology (DST), Govt. of India and Indo-US Science and Technology Forum (IUSSTF) for the WISTEMM fellowship and Dept. of Physics, UC Davis where a part of this work was carried out. GCA, BK and DKS acknowledge BRICS grant DST/IMRCD/BRICS/PilotCall1/MuMeSTU/2017(G) for the present work. The work of SQW is supported by National Natural Science Foundation of China (Grants 11963001, 11533003, 11603006, 11673006, 11851304, and U1731239).
		

\bibliography{refag}{}

\begin{thebibliography}{}
\expandafter\ifx\csname natexlab\endcsname\relax\def\natexlab#1{#1}\fi
\providecommand{\url}[1]{\href{#1}{#1}}
\providecommand{\dodoi}[1]{doi:~\href{http://doi.org/#1}{\nolinkurl{#1}}}
\providecommand{\doeprint}[1]{\href{http://ascl.net/#1}{\nolinkurl{http://ascl.net/#1}}}
\providecommand{\doarXiv}[1]{\href{https://arxiv.org/abs/#1}{\nolinkurl{https://arxiv.org/abs/#1}}}

\bibitem[{{Andrews} \& {Smith}(2018)}]{2018MNRAS.477...74A}
{Andrews}, J.~E., \& {Smith}, N. 2018, \mnras, 477, 74,
  \dodoi{10.1093/mnras/sty584}

\bibitem[{{Arnett}(1980)}]{1980ApJ...237..541A}
{Arnett}, W.~D. 1980, \apj, 237, 541, \dodoi{10.1086/157898}

\bibitem[{{Arnett}(1982)}]{1982ApJ...253..785A}
---. 1982, \apj, 253, 785, \dodoi{10.1086/159681}

\bibitem[{{Becker}(2015)}]{2015ascl.soft04004B}
{Becker}, A. 2015, {HOTPANTS: High Order Transform of PSF ANd Template
  Subtraction}.
\newblock \doeprint{1504.004}

\bibitem[{{Chatzopoulos} {et~al.}(2012){Chatzopoulos}, {Wheeler}, \&
  {Vinko}}]{2012ApJ...746..121C}
{Chatzopoulos}, E., {Wheeler}, J.~C., \& {Vinko}, J. 2012, \apj, 746, 121,
  \dodoi{10.1088/0004-637X/746/2/121}

\bibitem[{{Chevalier}(1982)}]{1982ApJ...258..790C}
{Chevalier}, R.~A. 1982, \apj, 258, 790, \dodoi{10.1086/160126}

\bibitem[{{Chevalier} \& {Fransson}(1994)}]{1994ApJ...420..268C}
{Chevalier}, R.~A., \& {Fransson}, C. 1994, \apj, 420, 268,
  \dodoi{10.1086/173557}

\bibitem[{{Chugai} \& {Danziger}(1994)}]{1994MNRAS.268..173C}
{Chugai}, N.~N., \& {Danziger}, I.~J. 1994, \mnras, 268, 173,
  \dodoi{10.1093/mnras/268.1.173}

\bibitem[{{Cooke} {et~al.}(2010){Cooke}, {Ellis}, {Nugent}, {Howell},
  {Sullivan}, \& {Gal-Yam}}]{2010ATel.2491....1C}
{Cooke}, J., {Ellis}, R.~S., {Nugent}, P.~E., {et~al.} 2010, The Astronomer's
  Telegram, 2491, 1

\bibitem[{{Crowther} \& {Smartt}(2007)}]{2007A&G....48a..35C}
{Crowther}, P., \& {Smartt}, S. 2007, Astronomy and Geophysics, 48, 1.35,
  \dodoi{10.1111/j.1468-4004.2007.48135.x}

\bibitem[{{Fassia} {et~al.}(2001){Fassia}, {Meikle}, {Chugai}, {Geballe},
  {Lundqvist}, {Walton}, {Pollacco}, {Veilleux}, {Wright}, {Pettini}, {Kerr},
  {Puchnarewicz}, {Puxley}, {Irwin}, {Packham}, {Smartt}, \&
  {Harmer}}]{2001MNRAS.325..907F}
{Fassia}, A., {Meikle}, W.~P.~S., {Chugai}, N., {et~al.} 2001, \mnras, 325,
  907, \dodoi{10.1046/j.1365-8711.2001.04282.x}

\bibitem[{{Foley} {et~al.}(2007){Foley}, {Smith}, {Ganeshalingam}, {Li},
  {Chornock}, \& {Filippenko}}]{2007ApJ...657L.105F}
{Foley}, R.~J., {Smith}, N., {Ganeshalingam}, M., {et~al.} 2007, \apjl, 657,
  L105, \dodoi{10.1086/513145}

\bibitem[{{Fremling} {et~al.}(2019){Fremling}, {Dugas}, \&
  {Sharma}}]{2019TNSCR.188....1F}
{Fremling}, C., {Dugas}, A., \& {Sharma}, Y. 2019, Transient Name Server
  Classification Report, 2019-188, 1

\bibitem[{{Gal-Yam}(2014)}]{2014AAS...22323502G}
{Gal-Yam}, A. 2014, in American Astronomical Society Meeting Abstracts, Vol.
  223, American Astronomical Society Meeting Abstracts \#223, 235.02

\bibitem[{{Gal-Yam} {et~al.}(2014){Gal-Yam}, {Arcavi}, {Ofek}, {Ben-Ami},
  {Cenko}, {Kasliwal}, {Cao}, {Yaron}, {Tal}, {Silverman}, {Horesh}, {De Cia},
  {Taddia}, {Sollerman}, {Perley}, {Vreeswijk}, {Kulkarni}, {Nugent},
  {Filippenko}, \& {Wheeler}}]{2014Natur.509..471G}
{Gal-Yam}, A., {Arcavi}, I., {Ofek}, E.~O., {et~al.} 2014, \nat, 509, 471,
  \dodoi{10.1038/nature13304}

\bibitem[{{Ginzburg} \& {Balberg}(2012)}]{2012ApJ...757..178G}
{Ginzburg}, S., \& {Balberg}, S. 2012, \apj, 757, 178,
  \dodoi{10.1088/0004-637X/757/2/178}

\bibitem[{{Gorbikov} {et~al.}(2014){Gorbikov}, {Gal-Yam}, {Ofek}, {Vreeswijk},
  {Nugent}, {Chotard}, {Kulkarni}, {Cao}, {De Cia}, {Yaron}, {Tal}, {Arcavi},
  {Kasliwal}, {Cenko}, {Sullivan}, \& {Chen}}]{2014MNRAS.443..671G}
{Gorbikov}, E., {Gal-Yam}, A., {Ofek}, E.~O., {et~al.} 2014, \mnras, 443, 671,
  \dodoi{10.1093/mnras/stu1184}

\bibitem[{{Hakobyan} {et~al.}(2012){Hakobyan}, {Adibekyan}, {Aramyan},
  {Petrosian}, {Gomes}, {Mamon}, {Kunth}, \& {Turatto}}]{2012A&A...544A..81H}
{Hakobyan}, A.~A., {Adibekyan}, V.~Z., {Aramyan}, L.~S., {et~al.} 2012, \aap,
  544, A81, \dodoi{10.1051/0004-6361/201219541}

\bibitem[{{Hosseinzadeh} {et~al.}(2019){Hosseinzadeh}, {McCully}, {Zabludoff},
  {Arcavi}, {French}, {Howell}, {Berger}, \& {Hiramatsu}}]{2019ApJ...871L...9H}
{Hosseinzadeh}, G., {McCully}, C., {Zabludoff}, A.~I., {et~al.} 2019, \apjl,
  871, L9, \dodoi{10.3847/2041-8213/aafc61}

\bibitem[{{Hosseinzadeh} {et~al.}(2017){Hosseinzadeh}, {Arcavi}, {Valenti},
  {McCully}, {Howell}, {Johansson}, {Sollerman}, {Pastorello}, {Benetti}, \&
  {Cao}}]{2017ApJ...836..158H}
{Hosseinzadeh}, G., {Arcavi}, I., {Valenti}, S., {et~al.} 2017, \apj, 836, 158,
  \dodoi{10.3847/1538-4357/836/2/158}

\bibitem[{{Jordi} {et~al.}(2006){Jordi}, {Grebel}, \&
  {Ammon}}]{2006A&A...460..339J}
{Jordi}, K., {Grebel}, E.~K., \& {Ammon}, K. 2006, \aap, 460, 339,
  \dodoi{10.1051/0004-6361:20066082}

\bibitem[{{Karamehmetoglu} {et~al.}(2019){Karamehmetoglu}, {Fransson},
  {Sollerman}, {Tartaglia}, {Taddia}, {De}, {Fremling}, {Bagdasaryan},
  {Barbarino}, {Bellm}, {Dekaney}, {Dugas}, {Giomi}, {Goobar}, {Graham}, {Ho},
  {Laher}, {Masci}, {Neill}, {Perley}, {Riddle}, {Rusholme}, \&
  {Soumagnac}}]{2019arXiv191006016K}
{Karamehmetoglu}, E., {Fransson}, C., {Sollerman}, J., {et~al.} 2019, arXiv
  e-prints, arXiv:1910.06016.
\newblock \doarXiv{1910.06016}

\bibitem[{{Khazov} {et~al.}(2016){Khazov}, {Yaron}, {Gal-Yam}, {Manulis},
  {Rubin}, {Kulkarni}, {Arcavi}, {Kasliwal}, {Ofek}, {Cao}, {Perley},
  {Sollerman}, {Horesh}, {Sullivan}, {Filippenko}, {Nugent}, {Howell}, {Cenko},
  {Silverman}, {Ebeling}, {Taddia}, {Johansson}, {Laher}, {Surace},
  {Rebbapragada}, {Wozniak}, \& {Matheson}}]{2016ApJ...818....3K}
{Khazov}, D., {Yaron}, O., {Gal-Yam}, A., {et~al.} 2016, \apj, 818, 3,
  \dodoi{10.3847/0004-637X/818/1/3}

\bibitem[{{Liu} {et~al.}(2018){Liu}, {Wang}, {Wang}, \&
  {Dai}}]{2018ApJ...856...59L}
{Liu}, L.-D., {Wang}, L.-J., {Wang}, S.-Q., \& {Dai}, Z.-G. 2018, \apj, 856,
  59, \dodoi{10.3847/1538-4357/aab157}

\bibitem[{{Lusk} \& {Baron}(2017)}]{2017PASP..129d4202L}
{Lusk}, J.~A., \& {Baron}, E. 2017, \pasp, 129, 044202,
  \dodoi{10.1088/1538-3873/aa5e49}

\bibitem[{{Lyman} {et~al.}(2016){Lyman}, {Bersier}, {James}, {Mazzali},
  {Eldridge}, {Fraser}, \& {Pian}}]{2016MNRAS.457..328L}
{Lyman}, J.~D., {Bersier}, D., {James}, P.~A., {et~al.} 2016, \mnras, 457, 328,
  \dodoi{10.1093/mnras/stv2983}

\bibitem[{{Morozova} {et~al.}(2017){Morozova}, {Piro}, \&
  {Valenti}}]{2017ApJ...838...28M}
{Morozova}, V., {Piro}, A.~L., \& {Valenti}, S. 2017, \apj, 838, 28,
  \dodoi{10.3847/1538-4357/aa6251}

\bibitem[{{Munari} \& {Zwitter}(1997)}]{1997A&A...318..269M}
{Munari}, U., \& {Zwitter}, T. 1997, \aap, 318, 269

\bibitem[{{Nicholl}(2018)}]{2018RNAAS...2..230N}
{Nicholl}, M. 2018, Research Notes of the American Astronomical Society, 2,
  230, \dodoi{10.3847/2515-5172/aaf799}

\bibitem[{{Nyholm} {et~al.}(2017){Nyholm}, {Sollerman}, {Taddia}, {Fremling},
  {Moriya}, {Ofek}, {Gal-Yam}, {De Cia}, {Roy}, {Kasliwal}, {Cao}, {Nugent}, \&
  {Masci}}]{2017A&A...605A...6N}
{Nyholm}, A., {Sollerman}, J., {Taddia}, F., {et~al.} 2017, \aap, 605, A6,
  \dodoi{10.1051/0004-6361/201629906}

\bibitem[{{Pastorello} {et~al.}(2007){Pastorello}, {Smartt}, {Mattila},
  {Eldridge}, {Young}, {Itagaki}, {Yamaoka}, {Navasardyan}, {Valenti}, \&
  {Patat}}]{2007Natur.447..829P}
{Pastorello}, A., {Smartt}, S.~J., {Mattila}, S., {et~al.} 2007, \nat, 447,
  829, \dodoi{10.1038/nature05825}

\bibitem[{{Pastorello} {et~al.}(2008){Pastorello}, {Mattila}, {Zampieri},
  {Della Valle}, {Smartt}, {Valenti}, {Agnoletto}, {Benetti}, {Benn}, {Branch},
  {Cappellaro}, {Dennefeld}, {Eldridge}, {Gal-Yam}, {Harutyunyan}, {Hunter},
  {Kjeldsen}, {Lipkin}, {Mazzali}, {Milne}, {Navasardyan}, {Ofek}, {Pian},
  {Shemmer}, {Spiro}, {Stathakis}, {Taubenberger}, {Turatto}, \&
  {Yamaoka}}]{2008MNRAS.389..113P}
{Pastorello}, A., {Mattila}, S., {Zampieri}, L., {et~al.} 2008, \mnras, 389,
  113, \dodoi{10.1111/j.1365-2966.2008.13602.x}

\bibitem[{{Pastorello} {et~al.}(2015{\natexlab{a}}){Pastorello}, {Wyrzykowski},
  {Valenti}, {Prieto}, {Koz{\l}owski}, {Udalski}, {Elias-Rosa},
  {Morales-Garoffolo}, {Anderson}, \& {Benetti}}]{2015MNRAS.449.1941P}
{Pastorello}, A., {Wyrzykowski}, {\L}., {Valenti}, S., {et~al.}
  2015{\natexlab{a}}, \mnras, 449, 1941, \dodoi{10.1093/mnras/stu2621}

\bibitem[{{Pastorello} {et~al.}(2015{\natexlab{b}}){Pastorello}, {Hadjiyska},
  {Rabinowitz}, {Valenti}, {Turatto}, {Fasano}, {Benitez-Herrera}, {Baltay},
  {Benetti}, \& {Botticella}}]{2015MNRAS.449.1954P}
{Pastorello}, A., {Hadjiyska}, E., {Rabinowitz}, D., {et~al.}
  2015{\natexlab{b}}, \mnras, 449, 1954, \dodoi{10.1093/mnras/stv335}

\bibitem[{{Pastorello} {et~al.}(2015{\natexlab{c}}){Pastorello}, {Benetti},
  {Brown}, {Tsvetkov}, {Inserra}, {Taubenberger}, {Tomasella}, {Fraser},
  {Rich}, {Botticella}, {Bufano}, {Cappellaro}, {Ergon}, {Gorbovskoy},
  {Harutyunyan}, {Huang}, {Kotak}, {Lipunov}, {Magill}, {Miluzio}, {Morrell},
  {Ochner}, {Smartt}, {Sollerman}, {Spiro}, {Stritzinger}, {Turatto},
  {Valenti}, {Wang}, {Wright}, {Yurkov}, {Zampieri}, \&
  {Zhang}}]{2015MNRAS.449.1921P}
{Pastorello}, A., {Benetti}, S., {Brown}, P.~J., {et~al.} 2015{\natexlab{c}},
  \mnras, 449, 1921, \dodoi{10.1093/mnras/stu2745}

\bibitem[{{Pastorello} {et~al.}(2015{\natexlab{d}}){Pastorello}, {Prieto},
  {Elias-Rosa}, {Bersier}, {Hosseinzadeh}, {Morales-Garoffolo}, {Noebauer},
  {Taubenberger}, {Tomasella}, \& {Kochanek}}]{2015MNRAS.453.3649P}
{Pastorello}, A., {Prieto}, J.~L., {Elias-Rosa}, N., {et~al.}
  2015{\natexlab{d}}, \mnras, 453, 3649, \dodoi{10.1093/mnras/stv1812}

\bibitem[{{Pastorello} {et~al.}(2016){Pastorello}, {Wang}, {Ciabattari},
  {Bersier}, {Mazzali}, {Gao}, {Xu}, {Zhang}, {Tokuoka}, \&
  {Benetti}}]{2016MNRAS.456..853P}
{Pastorello}, A., {Wang}, X.~F., {Ciabattari}, F., {et~al.} 2016, \mnras, 456,
  853, \dodoi{10.1093/mnras/stv2634}

\bibitem[{{Poznanski} {et~al.}(2012){Poznanski}, {Prochaska}, \&
  {Bloom}}]{2012MNRAS.426.1465P}
{Poznanski}, D., {Prochaska}, J.~X., \& {Bloom}, J.~S. 2012, \mnras, 426, 1465,
  \dodoi{10.1111/j.1365-2966.2012.21796.x}

\bibitem[{{Prentice} {et~al.}(2016){Prentice}, {Mazzali}, {Pian}, {Gal-Yam},
  {Kulkarni}, {Rubin}, {Corsi}, {Fremling}, {Sollerman}, {Yaron}, {Arcavi},
  {Zheng}, {Kasliwal}, {Filippenko}, {Cenko}, {Cao}, \&
  {Nugent}}]{2016MNRAS.458.2973P}
{Prentice}, S.~J., {Mazzali}, P.~A., {Pian}, E., {et~al.} 2016, \mnras, 458,
  2973, \dodoi{10.1093/mnras/stw299}

\bibitem[{{Prentice} {et~al.}(2019){Prentice}, {Ashall}, {James}, {Short},
  {Mazzali}, {Bersier}, {Crowther}, {Barbarino}, {Chen}, {Copperwheat},
  {Darnley}, {Denneau}, {Elias-Rosa}, {Fraser}, {Galbany}, {Gal-Yam},
  {Harmanen}, {Howell}, {Hosseinzadeh}, {Inserra}, {Kankare}, {Karamehmetoglu},
  {Lamb}, {Limongi}, {Maguire}, {McCully}, {Olivares E}, {Piascik}, {Pignata},
  {Reichart}, {Rest}, {Reynolds}, {Rodr{\'\i}guez}, {Saario}, {Schulze},
  {Smartt}, {Smith}, {Sollerman}, {Stalder}, {Sullivan}, {Taddia}, {Valenti},
  {Vergani}, {Williams}, \& {Young}}]{2019MNRAS.485.1559P}
{Prentice}, S.~J., {Ashall}, C., {James}, P.~A., {et~al.} 2019, \mnras, 485,
  1559, \dodoi{10.1093/mnras/sty3399}

\bibitem[{{Sanders} {et~al.}(2013){Sanders}, {Soderberg}, {Foley}, {Chornock},
  {Milisavljevic}, {Margutti}, {Drout}, {Moe}, {Berger}, \&
  {Brown}}]{2013ApJ...769...39S}
{Sanders}, N.~E., {Soderberg}, A.~M., {Foley}, R.~J., {et~al.} 2013, \apj, 769,
  39, \dodoi{10.1088/0004-637X/769/1/39}

\bibitem[{{Schlafly} \& {Finkbeiner}(2011)}]{2011ApJ...737..103S}
{Schlafly}, E.~F., \& {Finkbeiner}, D.~P. 2011, \apj, 737, 103,
  \dodoi{10.1088/0004-637X/737/2/103}

\bibitem[{{Schlegel}(1990)}]{1990MNRAS.244..269S}
{Schlegel}, E.~M. 1990, \mnras, 244, 269

\bibitem[{{Shivvers} {et~al.}(2016){Shivvers}, {Zheng}, {Mauerhan}, {Kleiser},
  {Van Dyk}, {Silverman}, {Graham}, {Kelly}, {Filippenko}, \&
  {Kumar}}]{2016MNRAS.461.3057S}
{Shivvers}, I., {Zheng}, W.~K., {Mauerhan}, J., {et~al.} 2016, \mnras, 461,
  3057, \dodoi{10.1093/mnras/stw1528}

\bibitem[{{Silverman} {et~al.}(2010){Silverman}, {Kleiser}, {Morton}, \&
  {Filippenko}}]{2010CBET.2223....1S}
{Silverman}, J.~M., {Kleiser}, I.~K.~W., {Morton}, A.~J.~L., \& {Filippenko},
  A.~V. 2010, Central Bureau Electronic Telegrams, 2223, 1

\bibitem[{{Smith} {et~al.}(2008){Smith}, {Foley}, \&
  {Filippenko}}]{2008ApJ...680..568S}
{Smith}, N., {Foley}, R.~J., \& {Filippenko}, A.~V. 2008, \apj, 680, 568,
  \dodoi{10.1086/587860}

\bibitem[{{Stetson}(1987)}]{1987PASP...99..191S}
{Stetson}, P.~B. 1987, \pasp, 99, 191, \dodoi{10.1086/131977}

\bibitem[{{Sun} {et~al.}(2019){Sun}, {Maund}, {Hirai}, {Crowther}, \&
  {Podsiadlowski}}]{2019arXiv190907999S}
{Sun}, N.-C., {Maund}, J.~R., {Hirai}, R., {Crowther}, P.~A., \&
  {Podsiadlowski}, P. 2019, arXiv e-prints, arXiv:1909.07999.
\newblock \doarXiv{1909.07999}

\bibitem[{{Taddia} {et~al.}(2013){Taddia}, {Stritzinger}, {Sollerman},
  {Phillips}, {Anderson}, {Boldt}, {Campillay}, {Castell{\'o}n}, {Contreras},
  {Folatelli}, {Hamuy}, {Heinrich-Josties}, {Krzeminski}, {Morrell}, {Burns},
  {Freedman}, {Madore}, {Persson}, \& {Suntzeff}}]{2013A&A...555A..10T}
{Taddia}, F., {Stritzinger}, M.~D., {Sollerman}, J., {et~al.} 2013, \aap, 555,
  A10, \dodoi{10.1051/0004-6361/201321180}

\bibitem[{{Tody}(1986)}]{1986SPIE..627..733T}
{Tody}, D. 1986, Society of Photo-Optical Instrumentation Engineers (SPIE)
  Conference Series, Vol. 627, {The IRAF Data Reduction and Analysis System},
  ed. D.~L. {Crawford}, 733, \dodoi{10.1117/12.968154}

\bibitem[{{Tody}(1993)}]{1993ASPC...52..173T}
---. 1993, Astronomical Society of the Pacific Conference Series, Vol.~52,
  {IRAF in the Nineties}, ed. R.~J. {Hanisch}, R.~J.~V. {Brissenden}, \&
  J.~{Barnes}, 173

\bibitem[{{Valenti} {et~al.}(2011){Valenti}, {Fraser}, {Benetti}, {Pignata},
  {Sollerman}, {Inserra}, {Cappellaro}, {Pastorello}, {Smartt}, {Ergon},
  {Botticella}, {Brimacombe}, {Bufano}, {Crockett}, {Eder}, {Fugazza},
  {Haislip}, {Hamuy}, {Harutyunyan}, {Ivarsen}, {Kankare}, {Kotak}, {Lacluyze},
  {Magill}, {Mattila}, {Maza}, {Mazzali}, {Reichart}, {Taubenberger},
  {Turatto}, \& {Zampieri}}]{2011MNRAS.416.3138V}
{Valenti}, S., {Fraser}, M., {Benetti}, S., {et~al.} 2011, \mnras, 416, 3138,
  \dodoi{10.1111/j.1365-2966.2011.19262.x}

\bibitem[{{Valenti} {et~al.}(2014){Valenti}, {Sand}, {Pastorello}, {Graham},
  {Howell}, {Parrent}, {Tomasella}, {Ochner}, {Fraser}, {Benetti}, {Yuan},
  {Smartt}, {Maund}, {Arcavi}, {Gal-Yam}, {Inserra}, \&
  {Young}}]{2014MNRAS.438L.101V}
{Valenti}, S., {Sand}, D., {Pastorello}, A., {et~al.} 2014, \mnras, 438, L101,
  \dodoi{10.1093/mnrasl/slt171}

\bibitem[{{Valenti} {et~al.}(2016){Valenti}, {Howell}, {Stritzinger}, {Graham},
  {Hosseinzadeh}, {Arcavi}, {Bildsten}, {Jerkstrand}, {McCully}, {Pastorello},
  {Piro}, {Sand}, {Smartt}, {Terreran}, {Baltay}, {Benetti}, {Brown},
  {Filippenko}, {Fraser}, {Rabinowitz}, {Sullivan}, \&
  {Yuan}}]{2016MNRAS.459.3939V}
{Valenti}, S., {Howell}, D.~A., {Stritzinger}, M.~D., {et~al.} 2016, \mnras,
  459, 3939, \dodoi{10.1093/mnras/stw870}

\bibitem[{{Wang} \& {Li}(2019)}]{2019arXiv190512623W}
{Wang}, S.-Q., \& {Li}, L. 2019, arXiv e-prints, arXiv:1905.12623.
\newblock \doarXiv{1905.12623}

\bibitem[{{Zhang} {et~al.}(2019){Zhang}, {Xing}, \&
  {Wang}}]{2019ATel12410....1Z}
{Zhang}, J., {Xing}, L., \& {Wang}, X. 2019, The Astronomer's Telegram, 12410

\end{thebibliography}
\bibliographystyle{aasjournal}





\end{document}